\documentclass[conference]{IEEEtran}
\usepackage{amsmath,amssymb,bm}
\usepackage{array}
\usepackage{algorithm}
\usepackage{algorithmic}
\usepackage{booktabs}
\usepackage{cite}
\usepackage{graphicx}
\usepackage[caption=false,font=footnotesize]{subfig}
\usepackage[acronym]{glossaries}
\usepackage[table]{xcolor}

\newcommand{\E}{\mathbb{E}}

\newcommand{\tx}{\mathrm{tx}}

\newcolumntype{C}[1]{>{\centering\arraybackslash}p{#1}}
\newcommand{\tabletextsize}{\scriptsize}

\newacronym{isac}{ISAC}{Integrated Sensing and Communication}
\newacronym{an}{AN}{Artificial Noise}
\newacronym{awgn}{AWGN}{Additive White Gaussian Noise}
\newacronym{ber}{BER}{Bit Error Rate}
\newacronym{ldpc}{LDPC}{Low-Density Parity-Check}
\newacronym{qpsk}{QPSK}{Quadrature Phase-Shift Keying}
\newacronym{ofdm}{OFDM}{Orthogonal Frequency-Division Multiplexing}
\newacronym{rf}{RF}{Radio Frequency}
\newacronym{5gnr}{5G NR}{Fifth-generation New Radio}
\newacronym{uav}{UAV}{Unmanned Aerial Vehicle}
\newacronym{los}{LOS}{Line-of-Sight}
\newacronym{mcs}{MCS}{Modulation and Coding Scheme}
\newacronym{bs}{BS}{Base Station}
\newacronym{snr}{SNR}{Signal-to-Noise Ratio}
\newacronym{psr}{PSR}{Peak Sidelobe Ratio}
\newacronym{crb}{CRB}{Cramer-Rao Bound}
\newacronym{dft}{DFT}{Discrete Fourier Transform}
\newacronym{ser}{SER}{Symbol Error Rate}
\newacronym{csprng}{CSPRNG}{Cryptographically Secure Pseudorandom Number Generator}
\newacronym{gf}{GF}{Gram Feature}
\newacronym{mae}{MAE}{mean absolute error}
\newacronym{mc}{MC}{Monte Carlo}
\newacronym{pls}{PLS}{Physical-Layer Security}
\newacronym{cnn}{CNN}{Convolutional Neural Network}
\newacronym{cfar}{CFAR}{Constant False-Alarm Rate}
\newacronym{std}{STD}{standard deviation}
\newacronym{em}{EM}{expectation-maximization}
\newacronym{nr}{NR}{New Radio}
\newacronym{ifft}{IFFT}{inverse fast Fourier transform}
\newacronym{rms}{RMS}{root mean square}
\newacronym{iot}{IoT}{Internet of Things}
\newacronym{gelu}{GELU}{Gaussian Error Linear Unit}

\title{Waveform Randomization for Secure ISAC Against Informed Eavesdroppers}

\author{
\IEEEauthorblockN{Zexin~Fang\IEEEauthorrefmark{1},~Bin~Han\IEEEauthorrefmark{1},
Shuangyang~Li\IEEEauthorrefmark{2},
and~Hans~D.~Schotten\IEEEauthorrefmark{1}\IEEEauthorrefmark{3}}
\IEEEauthorblockA{\IEEEauthorrefmark{1}{RPTU University Kaiserslautern-Landau},
\IEEEauthorrefmark{2}{Technische Universit\"at Berlin},\\
\IEEEauthorrefmark{3}{German Research Center for Artificial Intelligence (DFKI), Germany}}
}

\begin{document}
\maketitle

\begin{abstract}
We study waveform-level protection of \gls{isac} against a
\glsentryshort{uav}-mounted eavesdropper (Eve) that knows the waveform
construction and the phase-key generation rules. The novelty of this work
is a systematic, budget-explicit attack analysis: instead of relying on
undisclosed design details, we quantify what an informed Eve can learn and
reconstruct, and how much computation this takes. We protect the sensing
reference with Fourier-curve modulation, phase keys, and tangent \gls{an},
with burst-specific information shared over secure backhaul. We compare
three key-generation rules against offline \gls{cnn}-based sequence
identification and online Adam and phase-aware \gls{em} key
recovery, under complete or partial nominal-sequence knowledge. A
power-domain model couples residual key error with imperfect \gls{an}
removal and predicts coherent delay locking. Results show that reusable key
structure permits sequence identification even when demodulation fails,
position-dependent keying raises recovery difficulty, and seed-expanded
keying resists the evaluated classifier within the learning budget.
Waveform simulations under multipath, clutter, and echo accumulation
confirm the sensing impact.
\end{abstract}
\glsresetall


\section{Introduction}

\gls{isac} integrates data transmission and environmental sensing through
shared resources and hardware, offering improved spectral efficiency
and reduced deployment cost for next-generation
networks~\cite{liu2020isac}. However, the same transmitted signals that
support legitimate sensing can also expose environmental information to
unintended receivers~\cite{qu2024privacy,wang2025generative}. Many
\gls{isac} systems reuse pilots or other known reference signals for sensing.
These signals are accessible to authorized communication users and may also
be intercepted by external eavesdroppers outside the system. This can
further enable either receiver to perform unauthorized
detection, localization, or tracking~\cite{zou2024securing}.

A further threat is passive direct--echo bypass sensing: Eve correlates
an intercepted direct-path signal with target reflections without knowing
the signal structure.
Such sensing with non-cooperative illuminators is well established in
passive radar, including performance analysis and experimental detection
and tracking~\cite{griffiths2005pcl,howland2005fm}.
However, its performance is highly sensitive to reference-channel quality and
interference in the surveillance channel~\cite{liu2015passivecc}.
In urban environments, clutter and background radio interference from
concurrent transmissions can further degrade echo detection. For a moving receiver, channel imbalances
also complicate direct-path and clutter suppression~\cite{blasone2020passivedpca},
while sampling-clock mismatch degrades reference alignment~\cite{searle2021clock}.
Direct--echo bypass sensing is well established, but its performance depends
on reference quality and surveillance-channel interference.
We instead focus on less-explored informed coherent sensing attacks, where
Eve exploits nominal-reference knowledge to reconstruct a coherent reference.

Sensing-security research adapts
established \gls{pls} tools, including beamforming and
\gls{an}~\cite{goel2008artificial,matsumine2025survey}. These designs aim to limit
the target and environmental information available
to unauthorized receivers while preserving legitimate sensing performance.
Zou et al.~\cite{zou2024securing} consider
a communication user acting as a sensing Eve and design beamforming with
\gls{an} to limit its sensing mutual information while preserving legitimate
sensing and communication quality. Ren et al.~\cite{ren2024cellfree}
extend joint protection to cell-free \gls{isac}, optimizing cooperative
information and sensing beamforming against sensing eavesdroppers.
Joint signal perturbation and beamforming also
target sensing-signal indistinguishability~\cite{huang2026isacprivacy}.
Papanikolaou et al.~\cite{papanikolaou2026privacy} design robust transmit
covariances to misdirect unauthorized angular sensing under Eve-angle
uncertainty.
However, spatial protection often requires Eve's channel or location
information, or a specified uncertainty region. Obtaining and maintaining
this information is challenging for a passive, mobile Eve.
These spatial and covariance-domain designs motivate
waveform-level protection when spatial separation is unfavorable.
A \gls{uav}-mounted Eve can reposition to seek a strong \gls{los} link
and intercept energy intended for legitimate nodes; protection should
therefore be assessed under favorable interception geometry.

Waveform-domain methods modify the signal available to unauthorized
sensing receivers and can provide protection without knowledge of Eve's
geometry or reliance on spatial separation. Han et al.~\cite{han2026sensingsecure} engineer the
ambiguity function to introduce artificial targets and increase Eve's range
ambiguity. Du et al.~\cite{du2026kld} develop a detection-oriented
Kullback--Leibler-divergence metric accounting for mainlobe, sidelobe, and
noise contributions, and optimize time--frequency amplitude and phase
perturbations. Their analysis connects waveform shaping to the sensing-security
and communication tradeoff. Erdem et al.~\cite{erdem2026secure} introduce
artificial phase and Doppler shifts for passive \gls{ofdm} sensing
protection and demonstrate cooperative compensation through measurements.
Temiz et al.~\cite{temiz2026fmcw} combine index modulation and phase coding
in frequency-modulated continuous-wave chirps to impair unauthorized
detection and ranging. Zhang et al.~\cite{zhang2026ambsentry} instead use
ambient \gls{iot} devices as cooperative jammers and ghost targets, jointly
designing transmit beamforming and device reflection modulations.
Keskin et al.~\cite{keskin2026phasenoise} study
oscillator phase noise, showing that different phase-noise correlations in
monostatic and bistatic reception create a tradeoff among legitimate ranging,
unauthorized ranging, and communication performance. Unlike a
generated phase key, this mechanism exploits an oscillator-induced asymmetry.
These approaches illustrate that unauthorized detection and estimation can
be impaired through waveform manipulation rather than spatial suppression
alone.

However, waveform-domain protection faces two threats from an informed Eve.
First, Eve may estimate and compensate for the waveform modifications.
In \gls{isac}, one countermeasure strategy is to disguise the sensing
reference among communication signals.
Phase keys and \gls{an} in~\cite{han2026matched} can prevent correct
demodulation by Eve, thereby hindering her identification of similar
underlying bit-sequence structures in the protected signals for coherent
sensing. Second, with knowledge of the security design and
reference-signal generation, Eve may bypass the protection using a
specifically trained machine-learning model. This risk remains
underexplored and increasingly relevant.
Existing waveform-domain protections~\cite{han2026sensingsecure,du2026kld,erdem2026secure,temiz2026fmcw,zhang2026ambsentry,keskin2026phasenoise}
are evaluated against an Eve who lacks the protection parameters, or, as
in~\cite{erdem2026secure}, against a single heuristic phase-tracking Eve in
one measured scenario. Security should therefore not rely on undisclosed
design details or estimation complexity alone. We analyze what Eve can
learn and reconstruct when waveform construction and key-generation rules
are available; secrecy rests only on the burst-specific key shared over
secure backhaul.

We consider an \gls{ofdm} \gls{isac} system facing a \gls{uav}-mounted Eve
with favorable interception geometry. The system can switch modulation
orders, using \gls{qpsk} for lower-rate critical communication, such as
signaling, and sensing, and higher-order modulation for
high-throughput communication, consistent with \gls{nr} modulation
flexibility~\cite{3gpp38211,3gpp38214}. \textcolor{black}{Coordinated \glspl{bs} share the sensing
schedule, nominal-sequence seed, and keys over secure
backhaul, enabling reference reconstruction and coherent sensing.}
Communication-domain and sensing-domain keys are separated to prevent
communication access from exposing the sensing reference.
This article extends our conference work on phase-keyed
Fourier-curve waveform protection~\cite{fang2026waveform}. Its contributions are:

\begin{itemize}
\item \textbf{Protection framework.} We combine Fourier-curve modulation,
tangent \gls{an}, and phase keying in a two-layer design that conceals
sequence identity and protects coherent-reference reconstruction while
preserving legitimate communication and sensing. This design hinders
sequence identification and obstructs direct
key extraction when the nominal sequence is exposed.
\item \textbf{Attack-aware key analysis.} We evaluate three key-generation
rules (shared, smooth position-dependent, seed-expanded) against an
escalating set of informed-Eve attacks: blind cross-burst correlation,
offline \gls{cnn}-based root identification, and online Adam and phase-aware
\gls{em} key recovery. We report recovery accuracy and computational cost under
complete and partial nominal-sequence knowledge.
\item \textbf{Leakage-to-sensing analysis.} We expose \gls{qpsk}-specific
harmonic leakage, derive the symbol-agreement probability for sequence
localization, and develop a power-domain model that couples residual key
error with imperfect \gls{an} removal and predicts correct-delay locking,
validated by waveform simulations under multipath, clutter, and echo
accumulation.
\end{itemize}

\textbf{Compared with the conference version~\cite{fang2026waveform}},
which introduced phase-keyed Fourier-curve waveform protection, this article
makes the following additions:
\begin{enumerate}
\item It replaces a single key model with a comparison of shared, smooth
position-dependent, and seed-expanded keys under \gls{cnn}-based sequence
identification.
\item It adds online key-recovery attacks (Adam and phase-aware \gls{em}) with
complete or partial nominal-sequence knowledge, with and without
harmonic-structure initialization, and compares detection and key-recovery
accuracy across computational budgets.
\item It derives the symbol-agreement probability from the \gls{ser} to
characterize sequence localization, and adds a power-domain model linking
phase-key mismatch to residual tangent \gls{an} and coherent delay locking,
evaluated at waveform level by comparing a coordinated legitimate receiver
with Eve using imperfect key and \gls{an} estimates.
\end{enumerate}

The remainder of this paper is organized as follows. Section~II presents
the secure communication and sensing model. Section~III introduces the
threat model, key-generation rules, inference attacks, and detection
metrics. Section~IV reports the numerical results, and Section~V concludes
the paper.

\section{Secure Communication and Sensing Model}

\subsection{Fourier-Curve Modulator}
Let $M=2^q$. Every group of $q$ coded bits selects an index
$i\in\{0,\ldots,M-1\}$, mapped to a phase
\begin{equation}
\theta_i=\frac{2\pi i}{M}.
\end{equation}
Rather than modulating a single phase as in conventional $M$-PSK, we use
$\theta_i$ to generate a $k$-dimensional codeword whose components are
harmonics of the same angle,
\begin{equation}
\bm{x}_i=\frac{1}{\sqrt{k}}
\begin{bmatrix}
        e^{j\theta_i} & e^{j2\theta_i} & \cdots & e^{jk\theta_i}
\end{bmatrix}^{\mathsf T},
\qquad \|\bm{x}_i\|=1.
\label{eq:curve_codeword}
\end{equation}
For $k>M$, the harmonic phases $m\theta_i$ wrap modulo $2\pi$ and repeat
cyclically across the components.
The $k$ components are determined by $i$, and may
be mapped to parallel subcarriers, orthogonal signaling dimensions, or
successive resource elements depending on the physical implementation.

Differentiating $\bm{x}_i$ with respect to $\theta_i$ traces the local
direction of the codeword curve as $i$ varies,
\begin{equation}
\bm{x}'(\theta_i)=\frac{j}{\sqrt{k}}
\begin{bmatrix}
        e^{j\theta_i} & 2e^{j2\theta_i} & \cdots &
        ke^{jk\theta_i}
\end{bmatrix}^{\mathsf T}.
\end{equation}
Normalizing by its norm,
\begin{equation}
    v_k=\sqrt{\frac{1}{k}\sum_{\ell=1}^{k}\ell^2},
\end{equation}
gives the unit tangent vector at $\bm{x}_i$,
\begin{equation}
\hat{\bm{t}}_i=\frac{\bm{x}'(\theta_i)}{v_k},
\qquad \|\hat{\bm{t}}_i\|=1.
\label{eq:unit_tangent}
\end{equation}
The tangent is orthogonal to the codeword itself,
$\Re\{\bm{x}_i^{\mathsf H}\hat{\bm{t}}_i\}=0$, in the real embedding
$\mathbb{R}^{2k}$.
\subsection{Tangent \glsentryshort{an} and Channel}
Before \gls{an} injection, the transmitter applies a shared phase-key vector
\(
\bm{\varphi}=[\varphi_1,\ldots,\varphi_k]^{\mathsf T}
\)
across the $k$ harmonic components, giving the keyed Fourier-curve codeword
\begin{equation}
\bm{x}_i(\bm{\varphi})
    =\frac{1}{\sqrt{k}}
\begin{bmatrix}
        e^{j(\theta_i+\varphi_1)} &
        e^{j(2\theta_i+\varphi_2)} &
\cdots &
        e^{j(k\theta_i+\varphi_k)}
\end{bmatrix}^{\mathsf T}.
\label{eq:keyed_codeword}
\end{equation}
The keyed tangent $\hat{\bm{t}}_i(\bm{\varphi})$ follows by differentiating
$\bm{x}_i(\bm{\varphi})$ with respect to $\theta_i$. The legitimate receiver
knows $\bm{\varphi}$ and builds the keyed codebook directly; an
unintended receiver without the key faces a geometry mismatch even before
\gls{an} is considered.

For codeword $r$, the transmitter draws a real
$\xi_r\sim\mathcal N(0,1)$, shared across $k$ harmonic components, and
forms
\begin{equation}
\bm{x}^{\tx}_{i,r}
    =\sqrt{1-\beta}\,\bm{x}_i(\bm{\varphi})
    +\sqrt{\beta}\,\xi_r\hat{\bm{t}}_i(\bm{\varphi}),
\qquad 0\leq\beta<1.
\label{eq:protected_codeword}
\end{equation}
The tangent is deterministic for fixed $i$, while the
amplitude $\xi_r$ is drawn independently per
codeword, not per component. Since $\bm{x}_i(\bm{\varphi})$ and
$\hat{\bm{t}}_i(\bm{\varphi})$ are unit-norm and orthogonal,
\begin{equation}
\E\!\left[\|\bm{x}^{\tx}_{i,r}\|^2\right]=1.
\end{equation}

After synchronization and channel equalization, the real-valued received
codeword is
\begin{equation}
\bm{Y}_r=\sqrt{1-\beta}\,\bm{x}_i(\bm{\varphi})
    +\sqrt{\beta}\,\xi_r\hat{\bm{t}}_i(\bm{\varphi})+\bm{N}_r,
\label{eq:received_model}
\end{equation}
where $\bm{Y}_r,\bm{x}_i,\hat{\bm{t}}_i,\bm{N}_r\in\mathbb{R}^{2k}$ and
$\bm{N}_r\sim\mathcal N(\bm{0},\sigma_c^2\bm{I}_{2k})$, where
$\sigma_c^2$ is the noise variance per real component.

\subsection{\glsentryshort{qpsk} Sensing-Waveform Design}
\label{sec:waveform_design}
We consider \gls{qpsk} phase sequences
\begin{equation}
    s[n]=e^{j\theta_n},
\enspace
\theta_n\in\left\{0,\frac{\pi}{2},\pi,\frac{3\pi}{2}\right\},
\enspace n=0,\ldots,N-1.
\end{equation}
In the $k$-harmonic Fourier construction,
each sensing symbol $s[n]$ maps to the $k$-component subcarrier-weight
codeword
\begin{equation}
    \bm{u}_k[n]=\frac{1}{\sqrt{k}}
    \left[
    e^{j\theta_n},
    e^{j2\theta_n},
    \ldots,
    e^{jk\theta_n}
    \right]^{\mathsf T}.
\label{eq:qpsk_harmonic_waveform}
\end{equation}
An \gls{ifft} converts the active \gls{ofdm} subcarrier coefficients into a
time-domain sensing burst.
Sensing uses combined harmonic autocorrelation:
\begin{equation}
    R_k[\tau]
    =
    \sum_n \bm{u}_k[n]^{\mathsf H}\bm{u}_k[n-\tau].
\end{equation}
The design objective is
\begin{equation}
J_k(\bm{\theta})
=
\sum_{\tau\neq 0}|R_k[\tau]|^2
+
\lambda
\max_{\tau\neq 0}|R_k[\tau]|^2 ,
\label{eq:qpsk_objective}
\end{equation}
where $\bm{\theta}=[\theta_0,\ldots,\theta_{N-1}]^{\mathsf T}$ collects the
\gls{qpsk} phases of the sensing sequence. The first term penalizes total
sidelobe energy and the second the peak sidelobe, with $\lambda$ controlling
their tradeoff; we set $\lambda=10$ throughout to emphasize \gls{psr}. Since
the phase alphabet is discrete, the resulting optimization is combinatorial;
we solve it via discrete coordinate descent with random restarts, repeated per
harmonic order $k$ and summarized in Algorithm~\ref{alg:qpsk_waveform_design}.

\textcolor{black}{The nominal sensing sequence is phase-keyed and protected by
tangent \gls{an}, with the phase key
$\bm{\varphi}_{\mathrm{s}}$ shared over secure backhaul.} The cooperating
\glspl{bs} also share the sensing schedule, \gls{ofdm} frame timing, active
resources, and nominal-sequence seed.
Standard synchronization provides coarse alignment,
while the target-dependent propagation delay remains to be estimated through
matched sensing. After compensating known front-end and channel distortions,
the sensing waveform is reconstructed, with residual timing uncertainty
further reduced through echo accumulation. For communication, the \gls{bs} and communication user instead share a
communication-domain key $\bm{\varphi}_{\mathrm{c}}$ \cite{han2026matched} to build
the keyed codebook and tangent directions,
enabling covariance-aware demodulation without the \gls{an} realization
and protecting the sensing reference from unintended receivers.

\begin{algorithm}[t]
\scriptsize
\caption{\glsentryshort{qpsk} harmonic-waveform design for fixed $k$}
\label{alg:qpsk_waveform_design}
\begin{algorithmic}[1]
\STATE Choose $N$, $k$, restarts $R$, sweeps $S$; define $\mathcal{Q}=\{0,\pi/2,\pi,3\pi/2\}$ and objective $J_k(\bm{\theta})$ in \eqref{eq:qpsk_objective}.
\FOR{$r=1,\ldots,R$}
\STATE Initialize $\bm{\theta}^{(r)}$ (structured or random seed).
\FOR{$s=1,\ldots,S$}
\STATE \texttt{improved} $\leftarrow$ false.
\FOR{each chip $n$ (random order)}
\STATE Set $\theta_n\leftarrow\arg\min_{\theta\in\mathcal{Q}}J_k$, others fixed; if $J_k$ decreased, \texttt{improved} $\leftarrow$ true.
\ENDFOR
\IF{\texttt{improved} = false}
\STATE \textbf{break} (stop restart $r$).
\ENDIF
\ENDFOR
\STATE Store best sequence from restart $r$.
\ENDFOR
\STATE Return the best sequence overall (or all within tolerance of it).
\end{algorithmic}
\end{algorithm}
\subsection{Covariance-Aware Demodulation}

\subsubsection{Unknown-\glsentryshort{an} Maximum Likelihood Receiver}
Unlike the sensing case, where the waveform $\bm{x}$ is known, communication has no such prior: the candidate
$i$ is itself the unknown, so its tangent direction $\hat{\bm{t}}_i$ is
unknown until $i$ is decided. The receiver therefore marginalizes $\xi_r$
rather than estimating it, giving
\begin{equation}
\bm{Y}_r\mid i\sim\mathcal N\!\left(
\sqrt{1-\beta}\bm{x}_i,\bm{\Sigma}_i
\right),
\end{equation}
where
\begin{equation}
\bm{\Sigma}_i=\sigma_c^2\bm{I}_{2k}
    +\beta\hat{\bm{t}}_i\hat{\bm{t}}_i^{\mathsf T}.
\label{eq:covariance}
\end{equation}
The \gls{an} increases variance only along candidate $i$'s tangent. Since all
tangents have unit norm,
\begin{equation}
\det(\bm{\Sigma}_i)
    =(\sigma_c^2)^{2k-1}(\sigma_c^2+\beta)
\end{equation}
is independent of $i$. Define $\bm{r}_i=\bm{Y}_r-\sqrt{1-\beta}\bm{x}_i$.
Applying the Sherman--Morrison identity yields the candidate metric
\begin{equation}
\Lambda_i(\bm{Y}_r)
    =\frac{\|\bm{r}_i\|^2}{\sigma_c^2}
    -\frac{\beta(\bm{r}_i^{\mathsf T}\hat{\bm{t}}_i)^2}
    {\sigma_c^2(\beta+\sigma_c^2)}.
\label{eq:ml_metric}
\end{equation}
The hard decision is $\hat{i}=\arg\min_i\Lambda_i$. The first term is
Euclidean; the second reduces the penalty for residuals along candidate
$i$'s \gls{an} direction.

\subsubsection{Soft Demapping and \glsentryshort{ldpc} Decoding}
Let $b_a(i)$ denote bit position $a\in\{0,\ldots,q-1\}$ in the binary label
of candidate $i$. The covariance-aware soft demapper computes
\begin{equation}
    L_{r,a}=\log\frac{
\sum_{i:b_a(i)=0}\exp[-\Lambda_i(\bm{Y}_r)/2]
    }{
\sum_{i:b_a(i)=1}\exp[-\Lambda_i(\bm{Y}_r)/2]
    }.
\label{eq:llr}
\end{equation}

The \gls{ldpc} code is applied only to the control and payload channels,
which must meet an explicit \gls{ber} requirement; the sensing waveform is not
encoded, since doing so would perturb the optimized phase sequence. Its
schedule and nominal sequence are instead shared over the secure backhaul
between the coordinated base stations. A standard rate-$1/2$ \gls{ldpc} code with
iterative belief-propagation decoding is adopted as a
representative coded layer for the control and payload channels.

\section{Threat Model}\label{sec:thmodel}

\subsection{Eavesdropping Detection Model}
Without being a communication user, an eavesdropper observing the direct link may
still infer coarse features such as occupied bandwidth, burst timing, and
active sensing subcarriers. Her cyclostationary detection can be hindered by
randomized burst timing.
We assume Eve (i) exploits
long-term observations to attempt cross-burst coherent reference
reconstruction, (ii) may know the waveform structure and key-generation
rules to identify the nominal sequence from received waveforms, and
(iii) may use nominal-sequence knowledge to recover phase keys
for coherent sensing.

\subsection{Known-Reference Correct-Lock Analysis}
For the coherent sensing attack, we first assume Eve has the exact complex
waveform reference $x[n]$.
We adopt a delay-only detection model, assuming Eve's residual Doppler
after compensation causes negligible inter-carrier interference and phase
drift over the coherent processing interval. We assume Eve is hovering or
loitering with an unobstructed view of the target and model
the reflected path as a single-tap \gls{los}-dominant channel.
These assumptions favor Eve and simplify the analysis; the correct-lock
upper bound below follows from ignoring bin competition. The received echo is
\begin{equation}\begin{split}
\mathcal{H}_0:&\quad r[n]=w[n],\\
\mathcal{H}_1:&\quad r[n]=g_r\,x[n-\ell_0]+w[n],
\end{split}
\label{eq:known_reference_hyp}
\end{equation}
where $\mathcal{H}_0$ denotes no target-generated
echo is present at Eve, $\mathcal{H}_1$ denotes the target-present
hypothesis, $g_r$ is the complex path gain, $\ell_0$ is the unknown echo
delay relative to the transmit reference, and
$w[n]\sim\mathcal{CN}(0,\sigma_w^2)$ is temporally independent \gls{awgn}.

Eve correlates the received echo with the known reference:
\begin{equation}
    z[\ell]=\sum_n r[n]\,x^*[n-\ell],
\qquad \ell=0,\dots,L-1,
\label{eq:matched_filter_delay}
\end{equation}
followed by the peak-search rule
\begin{equation}
\max_{\ell} |z[\ell]| \underset{\mathcal{H}_0}{\overset{\mathcal{H}_1}{\gtrless}} \eta,
\qquad
\hat\ell=\arg\max_\ell |z[\ell]|.
\label{eq:peak_search}
\end{equation}
\textcolor{black}{The threshold $\eta$ is calibrated under $\mathcal H_0$ to
satisfy the search-wide false-alarm probability
$P_{\rm FA}=\Pr(\max_\ell|z[\ell]|>\eta\mid\mathcal H_0)$.
We assume the true delay $\ell_0$ lies on the search grid but is unknown to
Eve. Correct locking requires both threshold crossing and selection of the
true bin.} Define the normalized autocorrelation of the sensing
burst,
\begin{equation}
\chi[\ell] \triangleq \frac{1}{E_s}\sum_n x[n]\,x^*[n-\ell],
\qquad E_s=\sum_n|x[n]|^2.
\label{eq:ambiguity_def}
\end{equation}

\subsubsection{Noise Model}
For a fully observed noise-free reference, complex Gaussian correlation noise has variance
\begin{equation}
\sigma_z^2 = \sigma_w^2 E_s,
\end{equation}
equal under both hypotheses, yielding
\begin{equation}\begin{split}
    &z[\ell_0]\mid \mathcal{H}_0 \sim \mathcal{CN}(0,\sigma_z^2),\\
    &z[\ell_0]\mid \mathcal{H}_1 \sim \mathcal{CN}(g_r E_s,\,\sigma_z^2).
\end{split}\label{eq:pd_statistic}
\end{equation}

\subsubsection{Correct-Bin and False-Lock Probabilities}
As before,
\begin{equation}\begin{split}
P_{\mathrm{corr}}&=\Pr\!\big(\hat{\ell}=\ell_0,\ |z[\hat{\ell}]|>\eta\big),\\
P_{\mathrm{false}}&=\Pr\!\big(\hat{\ell}\neq \ell_0,\ |z[\hat{\ell}]|>\eta\big),
\label{eq:pcorr_pfalse}\end{split}
\end{equation}
with $P_D=P_{\mathrm{corr}}+P_{\mathrm{false}}$ and
$P_{\mathrm{miss}}=1-P_D$. 

\subsubsection{Role of the Mainlobe}
{\color{black}Since $\chi[0]=1$, the true-bin mean is
$\mu_0=g_rE_s$. For the same calibrated threshold $\eta$,
$P_{\mathrm{corr}}\leq\Pr(|z[\ell_0]|>\eta)$, since a correct lock also
requires the true bin to win the search. Under this Gaussian model,
\begin{equation}
    P_{\mathrm{corr}} \;\leq\;
    Q_1\!\left(
\frac{\sqrt{2}|\mu_0|}{\sigma_z},
\frac{\sqrt{2}\eta}{\sigma_z}
\right),
\label{eq:pd_explicit}
\end{equation}
where $|\mu_0|^2/\sigma_z^2$ is the mainlobe \gls{snr}. The bound neglects
competition from other delay bins.}

    \subsubsection{Role of the Sidelobes}
$P_{\mathrm{false}}$ is governed by the
off-peak values $\chi[\ell]$, $\ell\neq0$: a large sidelobe at offset
$\ell'\neq0$ gives mean response $g_rE_s\chi[\ell']$ at bin
$\ell_0+\ell'$, which can cross $\eta$ and inflate $P_{\mathrm{false}}$, and
hence $P_D$, without improving ranging accuracy. A lower-sidelobe
waveform can exhibit \emph{lower} $P_D$ than a higher-sidelobe one while
achieving equal or higher $P_{\mathrm{corr}}$. \textcolor{black}{The peak sidelobe
ratio,}
$\mathrm{PSR}=1/\max_{\ell\neq0}|\chi[\ell]|^2$,
\textcolor{black}{together with the mainlobe \gls{snr} $|\mu_0|^2/\sigma_z^2$,
characterizes ranging reliability more completely than $P_D$ alone.}

\begin{figure*}[!t]
\centering
\includegraphics[width=0.697680\textwidth]{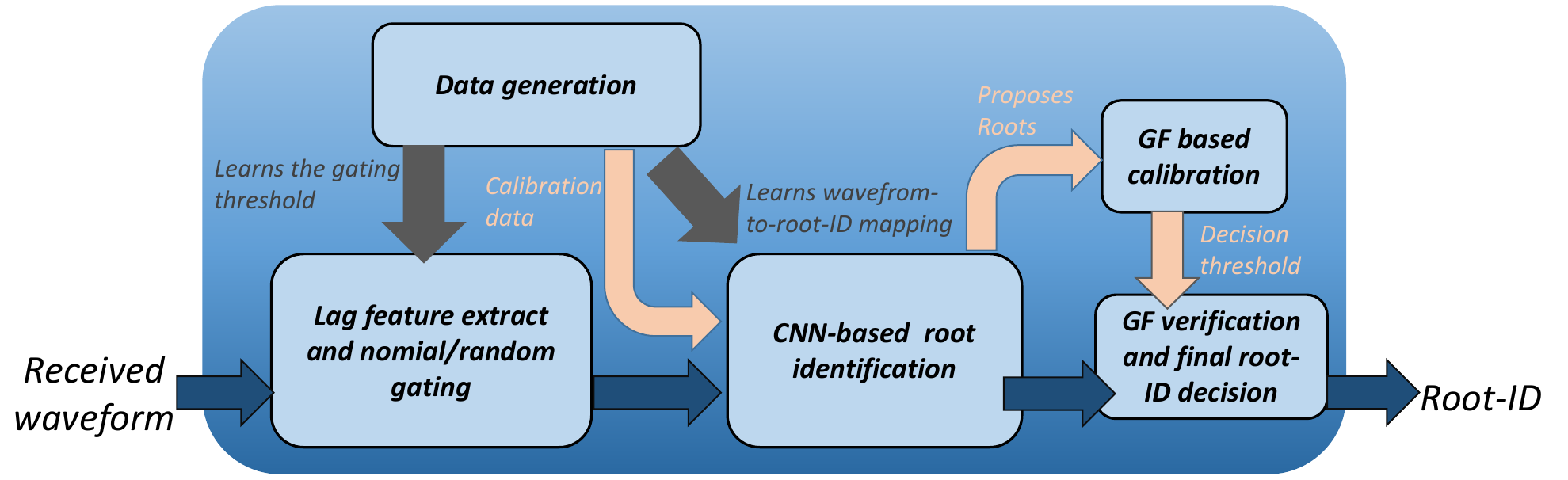}
\caption{\gls{cnn}-assisted sequence detection and root-identification pipeline.
The generated training data determine the nominal/random gate and the
waveform-to-root mapping. The \gls{cnn} proposes candidate roots, while the
independently calibrated \glsentryshort{gf} threshold verifies the final root-ID
decision.}
\label{fig:cnn_sequence_detection_pipeline}
\end{figure*}

\subsection{Phase-Key Generation Models}
We distinguish three burst-wise key-generation models.

\subsubsection{\textbf{\emph{Shared Lookup-Table Key}}}
The first model is the burst-wise extension of the phase-keyed Fourier-curve
lookup table in ~\cite{han2026matched}. For burst $b$, the
transmitter draws
\begin{equation}
\bm{\varphi}^{(b)}
=\left[\varphi_1^{(b)},\ldots,\varphi_k^{(b)}\right]^{\mathsf T},
\qquad
\varphi_m^{(b)}\overset{\mathrm{i.i.d.}}{\sim}\mathcal U[-\pi,\pi),
\label{eq:shared_burst_key}
\end{equation}
and applies this same $k$-coefficient vector to all $N$ symbols in that burst.
It therefore instantiates one burst-specific lookup table
$\{\bm{x}_i(\bm{\varphi}^{(b)}),
\hat{\bm{t}}_i(\bm{\varphi}^{(b)})\}_{i=0}^{M-1}$ and is refreshed every burst.

\subsubsection{\textbf{\emph{Smooth Position-Dependent Key}}}
The second model is introduced here as a structured benchmark, motivated by
controlled phase-perturbation methods for secure \gls{ofdm}
waveforms~\cite{erdem2026secure}.
For each burst, legitimate nodes use the shared key to select
$a_m$, $b_m$, and $c_m$ from predefined spaces for each harmonic
$m$, giving
\begin{equation}
\begin{aligned}
\varphi_m^{(b)}[n]
&=d_\varphi\!\left[
\sin\!\left(2\pi a_m n+c_m\right)\right.\\
&\qquad\left.
+0.55\sin\!\left(2\pi b_m n+0.7c_m\right)
\right],
\end{aligned}
\label{eq:smooth_burst_key}
\end{equation}
where $d_\varphi$ controls the phase-key depth; we set $d_\varphi=2.0$ in
the simulations. This equation produces a position-dependent keyed codebook
for each burst.

\subsubsection{\textbf{\emph{Seed-Expanded Pseudorandom Key}}}
The third model follows standard deterministic random-bit generation
\cite{nist_sp800_90a}. Before operation, the coordinated legitimate nodes
receive a shared master seed $K_{\rm seed}\in\{0,1\}^{\lambda}$ through the
secure backhaul and maintain a synchronized burst counter
$b$. The \gls{csprng} generates one pseudorandom value from this pair
for each position $n$ and harmonic $m$:
\begin{equation}
\begin{aligned}
\bm{U}^{(b)}&=\mathcal{G}(K_{\rm seed},b)\in[0,1)^{N\times k},\\
\bm{\Phi}^{(b)}&=2\pi\bm{U}^{(b)}-\pi\bm{1}_{N\times k},
\end{aligned}
\label{eq:seed_expanded_key}
\end{equation}
where $\Phi^{(b)}_{n,m}=\varphi_m^{(b)}[n]\in[-\pi,\pi)$;
$b$ separates burst domains.

With $q$-bit parameters, the \emph{shared} and \emph{smooth} rules require
$kq$ and $3kq$ bits per burst. The \emph{seed-expanded} rule shares a
$\lambda$-bit seed for local mask generation; its security depends on seed-space
size, at the cost of pseudorandom expansion.

\subsection{Eve Key- and Symbol-Recovery Models}
We assume an informed Eve knows the waveform construction in
Algorithm~\ref{alg:qpsk_waveform_design} and the phase-key generation rule.
The nominal-sequence class set is $\mathcal R=\{1,\ldots,R\}$, and the
seed space is $\mathcal S$. A seed $s\in\mathcal S$ selects a class through
$r=g(s)$. The seed is shared over secure backhaul and updated periodically.
Eve knows both spaces and the mapping, but not the realized seed or class. Eve
may use an offline-trained model to map the protected waveform
to its nominal root $r\in\mathcal R$, bypassing symbol and phase-key
estimation.

\subsubsection{\textbf{Offline Attack: \gls{cnn}-Based Sequence Detection and Root Identification}}
A data-driven Eve first separates nominal sensing bursts from random bursts
and then identifies the generating root of each accepted nominal burst, as
illustrated in Fig.~\ref{fig:cnn_sequence_detection_pipeline}.
The data-generation module knows the class set $\mathcal R$, the seed space
and selection rule, the phase-key generation rule, and the \gls{an} fraction
$\beta$. It uses this
 structural knowledge to generate labeled protected bursts. Each
training sample is generated by independently drawing a nominal root and a
phase-key realization from their known candidate spaces, while the
\gls{snr} is randomly selected from a predefined range.

{\color{black}Within the detection model, 
the nominal/random gate learns lag features from
the nominal and random data sets and either rejects the input or passes it to
the \gls{cnn}-based root-identification module. The \gls{cnn} then proposes candidate
roots, and the \gls{gf} module returns the final root-ID decision. Prior to
that, the
labeled calibration data comprise \emph{seen nominal} (nominal samples used to
train the \gls{cnn}), \emph{random} (random sequences of the same length), and
\emph{unseen nominal} (nominal samples not used in \gls{cnn} training). These samples
are fed to the \gls{cnn} to obtain candidate root IDs, from which the corresponding
\gls{gf} scores are calculated. The \gls{gf} calibration module then sets the
decision threshold from the score distributions of the three data sets.}

\textbf{Step 1: Lag-feature extraction and nominal/random gating.}
The input is the received active-subcarrier matrix
$\bm{Y}=[\bm y[0],\ldots,\bm y[N-1]]^{\mathsf T}
\in\mathbb{C}^{N\times k}$.
For each nonzero symbol lag, Eve computes the normalized correlation power
\begin{equation}
\rho_\ell=\left|\frac{\sum_{n=\ell}^{N-1}
\bm y[n]^{\mathsf H}\bm y[n-\ell]}
{\sum_{n=0}^{N-1}\|\bm y[n]\|^2}\right|^2,
\quad \ell=1,\ldots,N-1.
\label{eq:eve_lag_power}
\end{equation}
and forms the two-dimensional feature vector
\begin{equation}
\bm f=\left[-10\log_{10}\!\left(\max_\ell\rho_\ell\right),\;
2\sum_{\ell=1}^{N-1}\rho_\ell\right]^{\mathsf T}
\label{eq:eve_gate_features}
\end{equation}
whose entries are the received \gls{psr} and normalized two-sided sidelobe
energy, respectively. The feature vector $\bm f$ is standardized element-wise
using the training-set mean $\bm\mu_f$ and standard deviation $\bm\sigma_f$.
A linear logistic gate
produces
\begin{equation}
g(\bm f)=\bm w_g^{\mathsf T}
\left((\bm f-\bm\mu_f)\oslash\bm\sigma_f\right)+c_g,
\label{eq:eve_logistic_gate}
\end{equation}
where $\oslash$ denotes element-wise division,
$\bm w_g\in\mathbb{R}^{2}$ is the learned gate-weight vector, and
$c_g\in\mathbb{R}$ is the learned bias. For training sample $i$, let
$z_i=1$ denote a nominal burst and $z_i=0$ a random burst, and let
$p_i=\left(1+e^{-g(\bm f_i)}\right)^{-1}$. The parameters are learned using
mini-batch Adam to minimize
\begin{equation*}
(\hat{\bm w}_g,\hat c_g)=\arg\min_{\bm w_g,c_g}
-\frac{1}{B}\sum_{i=1}^{B}
\left[z_i\log p_i+(1-z_i)\log(1-p_i)\right].
\end{equation*}
Eve rejects the observation as random when $g(\bm f)<0$; otherwise, it
proceeds to root identification.

\textbf{Step 2: Phase-invariant \gls{cnn} root identification.}
Let
\begin{equation}
E_y=\frac{1}{N}\sum_{n=0}^{N-1}\|\bm y[n]\|^2.
\end{equation}
For $n\geq1$, Eve constructs the adjacent-symbol product
$\bm d[n]=\bm y[n]\odot\bm y[n-1]^*/E_y$, sets $\bm d[0]=\bm0$, and defines
\begin{equation}
\bm q[n]=\left[
\Re\{\bm d[n]\}^{\mathsf T},
\Im\{\bm d[n]\}^{\mathsf T},
\frac{\left(|\bm y[n]|^2\right)^{\mathsf T}}{E_y}
\right]^{\mathsf T}\in\mathbb{R}^{3k}.
\label{eq:eve_invariant_cnn_input}
\end{equation}
At each position, $\bm q[n]$ comprises real and imaginary parts of
the component-wise products between two consecutive $k$-component received
vectors and the normalized component powers of the current vector. These three
$k$-dimensional terms yield the $3k\times N$ \gls{cnn} input. The adjacent products
suppress phase components common to consecutive symbols, while the power terms
retain amplitude information. Table~\ref{tab:eve_cnn_architecture} summarizes
the \gls{cnn}, which uses \gls{gelu} activations and outputs root logits.

\begin{table}[H]
\centering
\caption{\gls{cnn} architecture for Eve-side nominal-root identification.}
\label{tab:eve_cnn_architecture}
\tabletextsize
\setlength{\tabcolsep}{3pt}
\renewcommand{\arraystretch}{1.05}
\begin{tabular}{lcc}
\toprule
\rowcolor{gray!20}
\textbf{Layer} & \textbf{Size / channels} & \textbf{Operation} \\
\midrule
\rowcolor{gray!8}
Input & $3k\times N$ & Phase-invariant features $\bm q[n]$ \\
\rowcolor{gray!15}
Conv. block 1 & $3k\to32$ & Conv1D, kernel 7, \glsentryshort{gelu} \\
\rowcolor{gray!8}
Conv. block 2 & $32\to32$ & Conv1D, kernel 7, \glsentryshort{gelu} \\
\rowcolor{gray!15}
Hidden layer & $32N\to64$ & Flatten, linear, \glsentryshort{gelu} \\
\rowcolor{gray!8}
Output & $64\to R$ & Nominal-root logits \\
\bottomrule
\end{tabular}
\end{table}

The root head minimizes multiclass cross-entropy over $B$ nominal bursts
with labels $r_i\in\mathcal R$:
\begin{equation}
\mathcal{L}_{\rm CNN}(\bm\psi)
=-\frac{1}{B}\sum_{i=1}^{B}
\log\frac{\exp(h_{i,r_i})}{\sum_{r=1}^{R}\exp(h_{i,r})},
\label{eq:eve_root_cnn_loss}
\end{equation}
where $\bm\psi$ denotes the \gls{cnn} parameters, $r\in\mathcal R$ indexes a
candidate root class, and $h_{i,r}$ is the corresponding logit for burst
$i$; hence, $h_{i,r_i}$ is the ground-truth-class logit. For an accepted burst, the \gls{cnn} returns root logits
$\bm h=[h_1,\ldots,h_R]^{\mathsf T}$ for the $R$ roots known during training.
Eve sorts these logits and retains the set $\mathcal C_L$ containing the
$L$ highest-ranked roots. We use $L=20$ to form a shortlist of candidate
roots.

\textbf{Step 3: \gls{gf} verification.}
The received burst is centered across symbol positions as
\begin{equation}
\tilde{\bm y}[n]=\bm y[n]-\frac{1}{N}\sum_{n'=0}^{N-1}\bm y[n'],
\end{equation}
where $n'$ is a dummy index over all $N$ received symbol positions.
The strict upper-triangular entries of the centered Gram matrix
$[\bm G_y]_{n,n'}=\tilde{\bm y}[n]^{\mathsf H}\tilde{\bm y}[n']$ form
$\bm g_y$. Using precomputed nominal-root vector $\bm g_r$, Eve scores
candidate $r$ by
\begin{equation}
s_r=\frac{\Re\{\bm g_r^{\mathsf H}\bm g_y\}}
{\|\bm g_r\|\,\|\bm g_y\|}.
\label{eq:eve_gram_similarity}
\end{equation}
The \gls{gf} module scores each shortlisted candidate.

\textbf{Step 4: Independent threshold calibration.}
The verification threshold is calibrated on an independent set as
\begin{equation}
\eta^*=\arg\max_\eta\frac{1}{2}\left[
\Pr(s^+\geq\eta)+\Pr(s^-<\eta)\right].
\label{eq:eve_verification_threshold}
\end{equation}
Here, $s_i^+=s_{i,r_i}$ is the \gls{gf} score assigned to the true root of
nominal calibration burst $i$. If that root is absent from $\mathcal C_L$,
the burst is counted as a missed identification. For random calibration burst
$j$, $s_j^-=\max_{r\in\mathcal C_L}s_{j,r}$ is the largest score among its
shortlisted candidates. Thus, the two probabilities measure correct-root
acceptance and random-burst rejection, respectively. Eq.~(\ref{eq:eve_verification_threshold})
balances these outcomes; the resulting $\eta^*$ is then fixed and used for
final decisions.

\textbf{Step 5: Final decision.}
Eve selects the shortlisted candidate with the largest \gls{gf} score,
$r^*=\arg\max_{r\in\mathcal C_L}s_r$, and returns $r^*$ as the detected
nominal-root ID only if
$s_{r^*}\geq\eta^*$. Otherwise, a burst that passed the logistic gate is
reported as an unknown nominal sequence. The attack returns a
root ID, an unknown-nominal decision, or random-burst rejection.

\subsubsection{\textbf{Online Attack: Fitting-Based Detection}}
This attack does not rely on the offline-trained \gls{cnn}.
\textcolor{black}{If the sensing
system reuses the nominal sequence associated with a fixed root $r\in\mathcal R$
for a long period, Eve may infer all or part
of it from the empirical symbol distribution at each position.} Given
this estimate, Eve fits the phase key and uses the fitting score to gate
whether a protected burst contains the nominal sequence. An accepted fit
reconstructs the complex reference for coherent matched filtering and delay
estimation; symbol-level localization provides only coarse timing.

Let $A_n\in\{0,1,2,3\}$ denote the nominal \gls{qpsk} label at symbol position
$n$, with $\theta_{A_n}=2\pi A_n/4$. The \gls{qpsk} harmonic construction has a
structural exposure. Since
$\theta_{A_n}\in\{0,\pi/2,\pi,3\pi/2\}$,
\begin{equation*}
e^{j4\theta_{A_n}}=1,
\qquad
e^{j5\theta_{A_n}}=e^{j\theta_{A_n}}.
\end{equation*}
Consequently, the fourth harmonic and the fifth-to-first relative phase can
remove the unknown symbol dependence and expose key-dependent structure.
These observations remain coupled to \gls{an} and do not reveal the complete
key, but can initialize or constrain the fitting algorithms.

The resulting attack is a model-informed constrained maximum-likelihood
problem. Let $\bm{\zeta}$ denote the unknown key parameters: $\bm{\varphi}$
for the \emph{shared key} in~\eqref{eq:shared_burst_key}, or
$\{a_m,b_m,c_m\}_{m=1}^{k}$ for the \emph{smooth key}
in~\eqref{eq:smooth_burst_key}. After treating the tangent-\gls{an}
coefficients $\bm{\xi}$ as nuisance parameters, Eve seeks
\begin{equation}
(\hat{\bm{\zeta}},\hat{\bm{A}}_{\mathcal U})
=\arg\max_{\bm{\zeta},\bm{A}_{\mathcal U}}
p\!\left(\bm{Y}\mid
\bm{A}_{\mathcal K},\bm{A}_{\mathcal U},\bm{\zeta}\right),
\label{eq:eve_joint_ml_fit}
\end{equation}
Here, $\bm{Y}$ is the received burst. Moreover, $\mathcal K$ and $\mathcal U$
are the known and unknown symbol-position
sets, respectively.
\textcolor{black}{The tangent-\gls{an} coefficients are projected out, and
unknown symbols fitted jointly with the key.} The non-convex likelihood
may admit equivalent decompositions.

\textbf{Adam-based continuous fitting:}
Adam is used as a first-order numerical solver
for this fitting problem~\cite{kingma2014adam,zhong2022rmocg}.
For candidate key parameters $\bm{\zeta}$ and a known symbol $A_n$, define
\begin{equation*}
\bm{a}_n(\bm{\zeta})
=\bm{y}[n]-\sqrt{1-\beta}\,
\bm{x}_{A_n}\!\left(\bm{\varphi}_{\bm{\zeta}}[n]\right),
\end{equation*}
where $\bm{\varphi}_{\bm{\zeta}}[n]$ is generated from the shared- or
smooth-key parameters. For a fixed candidate key, the per-symbol
tangent-\gls{an} coefficient has the least-squares estimate
\begin{equation*}
\hat{\xi}_n(\bm{\zeta})
 =\frac{\Re\!\left\{
 \hat{\bm{t}}_{A_n}^{\mathsf H}\!\left(\bm{\varphi}_{\bm{\zeta}}[n]\right)
 \bm{a}_n(\bm{\zeta})
 \right\}}
 {\sqrt{\beta}\,
 \left\|\hat{\bm{t}}_{A_n}\!\left(\bm{\varphi}_{\bm{\zeta}}[n]\right)\right\|^2}.
\end{equation*}
Eve substitutes this estimate into the residual-energy objective
\begin{equation}
\mathcal{L}_{\mathrm{Adam}}(\bm{\zeta})
 =
 \sum_{n\in\mathcal K}
 \left\|
 \bm{a}_n(\bm{\zeta})
 -\sqrt{\beta}\,\hat{\xi}_n(\bm{\zeta})
 \hat{\bm{t}}_{A_n}\!\left(\bm{\varphi}_{\bm{\zeta}}[n]\right)
 \right\|_2^2,
\label{eq:eve_adam_loss}
\end{equation}
where $\mathcal K$ contains the known symbol positions. The projection removes
the best-fitting tangent-\gls{an} component at each such position. The key
parameters are then updated
by Adam~\cite{kingma2014adam} according to the gradient of~\eqref{eq:eve_adam_loss}:
\begin{equation*}
\bm{\zeta}^{(q+1)}
 =\Pi_{\mathcal Z}\!\left(
 \bm{\zeta}^{(q)}-\alpha_q
 \frac{\widehat{\bm{m}}_q}{\sqrt{\widehat{\bm{v}}_q}+\epsilon}
 \right),
\end{equation*}
where $\widehat{\bm{m}}_q$ and $\widehat{\bm{v}}_q$ are Adam's bias-corrected
first- and second-moment estimates, $\alpha_q$ is the step size, and
$\epsilon$ avoids division by zero. $\Pi_{\mathcal Z}$ wraps phase parameters
and projects the remaining parameters onto their prescribed domains. If Eve
knows and exploits this structural exposure, the first restart is initialized
from the fourth-harmonic and fifth-to-first relative-phase observations.
Without this knowledge, all restarts are random. The procedure stops after a fixed iteration budget or
when the relative loss decrease falls below a prescribed tolerance.

\begin{algorithm}[t]
\scriptsize
\caption{Adam-based phase-key fitting by Eve}
\label{alg:eve_adam_key_fit}
\begin{algorithmic}[1]
\STATE Input received burst $\bm{y}$, candidate nominal sequence, $k$, $\beta$, iterations $Q$, and restarts $R$.
\FOR{$r=1,\ldots,R$}
\STATE If structural knowledge is available, use harmonic-informed initialization for the first restart; otherwise initialize randomly. Set Adam moments $\bm{m}^{(0)},\bm{v}^{(0)}$ to zero.
\FOR{$q=0,\ldots,Q-1$}
\STATE Generate $\bm{\varphi}_{\bm{\zeta}^{(q)}}[n]$ and form $\bm{x}_{A_n}$ and $\hat{\bm{t}}_{A_n}$ for every $n\in\mathcal K$.
\STATE Compute each $\hat{\xi}_n(\bm{\zeta}^{(q)})$ by tangent projection.
\STATE Evaluate $\mathcal{L}_{\mathrm{Adam}}$ and its gradient.
\STATE Update Adam moments and project $\bm{\zeta}^{(q+1)}$ onto $\mathcal Z$.
\ENDFOR
\STATE Store the key with the smallest final residual.
\ENDFOR
\STATE Return the best key and its per-symbol tangent-\gls{an} estimates.
\end{algorithmic}
\end{algorithm}

\textbf{Phase-aware \glsentryshort{em}:}
If the leaked nominal symbols are uncertain, Eve may use a phase-aware
\gls{em} procedure~\cite{dempster1977em,moon1996em},
following \gls{em}-based joint detection and estimation in \gls{ofdm} systems
\cite{choi2004emofdm}. Unlike Adam's fixed-label fit, phase-aware \gls{em}
evaluates all $M$ candidates at uncertain positions. For
$i\in\{0,\ldots,M-1\}$, define
\begin{equation*}
\bm{a}_{n,i}(\bm{\zeta})
=\bm{y}[n]-\sqrt{1-\beta}\,
\bm{x}_{i}\!\left(\bm{\varphi}_{\bm{\zeta}}[n]\right),
\end{equation*}
and profile out the per-symbol tangent-\gls{an} coefficient as
\begin{equation*}
\hat\xi_{n,i}(\bm{\zeta})
=\frac{\Re\!\left\{
\hat{\bm t}_{i}^{\mathsf H}\!\left(\bm\varphi_{\bm\zeta}[n]\right)
\bm a_{n,i}(\bm\zeta)\right\}}
{\sqrt{\beta}\,
\left\|\hat{\bm t}_{i}\!\left(\bm\varphi_{\bm\zeta}[n]\right)\right\|^2}.
\end{equation*}
The corresponding profile log score is
\begin{equation*}
\Lambda_{n,i}(\bm\zeta)
=-\frac{1}{2\sigma_c^2}
\left\|\bm a_{n,i}(\bm\zeta)
-\sqrt{\beta}\,\hat\xi_{n,i}(\bm\zeta)
\hat{\bm t}_{i}\!\left(\bm\varphi_{\bm\zeta}[n]\right)\right\|_2^2.
\end{equation*}
Let $\pi_{n,i}$ denote Eve's prior probability that position $n$ contains
symbol $i$. It is one-hot for a known label and uniform for an erased label.
At \gls{em} cycle $q$, the E-step computes
\begin{equation*}
q_{n,i}^{(q)}
=\frac{\pi_{n,i}\exp[\Lambda_{n,i}(\bm\zeta^{(q)})]}
{\sum_{j=0}^{M-1}\pi_{n,j}\exp[\Lambda_{n,j}(\bm\zeta^{(q)})]}.
\end{equation*}
Thus, the E-step uses the current key estimate to perform soft symbol
detection: $q_{n,i}^{(q)}$ is the posterior probability of candidate $i$ at
position $n$ after the candidate-specific \gls{an} coefficient has been
profiled out.
The M-step holds $q_{n,i}^{(q)}$ fixed and uses Adam to update
$\bm\zeta$ by minimizing
\begin{equation}
\mathcal{L}_{\mathrm{EM}}(\bm{\zeta})
=-\frac{1}{N}\sum_{n=0}^{N-1}\sum_{i=0}^{M-1}
q_{n,i}^{(q)}\Lambda_{n,i}(\bm{\zeta}),
\label{eq:eve_em_loss}
\end{equation}
After each update, $\bm\zeta$ is projected onto the key-parameter set
$\mathcal Z$. Thus, labels declared
known remain fixed, whereas erased or uncertain labels are refined jointly
with the key. The updated key changes the candidate scores used by the next
E-step, and the two steps alternate until the likelihood converges or the
iteration budget is exhausted. Multiple initializations are retained because
the likelihood is non-convex.

\begin{algorithm}[t]
\scriptsize
\caption{Phase-aware \glsentryshort{em} key estimation by Eve}
\label{alg:eve_em_key_fit}
\begin{algorithmic}[1]
\STATE Input received burst $\bm Y$, symbol priors $\pi_{n,i}$, $M$, $k$, $\beta$, $\sigma_c^2$, \glsentryshort{em} cycles $Q$, and restarts $R$.
\FOR{$a=1,\ldots,R$}
\STATE If structural knowledge is available, use harmonic-informed initialization for the first restart; otherwise initialize $\bm\zeta$ randomly in $\mathcal Z$.
\FOR{$q=0,\ldots,Q-1$}
\STATE Form $\bm{x}_i(\bm\varphi_{\bm\zeta^{(q)}}[n])$ and $\hat{\bm t}_i(\bm\varphi_{\bm\zeta^{(q)}}[n])$ for every $(n,i)$.
\STATE Profile out $\hat\xi_{n,i}$, evaluate $\Lambda_{n,i}$, and normalize it to obtain $q_{n,i}^{(q)}$.
\STATE Hold $q_{n,i}^{(q)}$ fixed and take Adam steps minimizing $\mathcal{L}_{\mathrm{EM}}$.
\STATE Project the updated $\bm\zeta^{(q+1)}$ onto $\mathcal Z$.
\STATE Stop if the relative likelihood change is below a tolerance.
\ENDFOR
\STATE Evaluate the marginal likelihood and store the best restart.
\ENDFOR
\STATE Return the best key and symbol estimates.
\end{algorithmic}
\end{algorithm}

\textbf{Fitting-score decision:}
Both algorithms can be used to detect the nominal waveform. For
$a\in\{\mathrm{Adam},\mathrm{EM}\}$, Eve computes
$d_a=\min_{\bm\zeta}\mathcal L_a(\bm\zeta)/c_a$, where
$c_{\rm Adam}=\|\bm Y\|_F^2$ and $c_{\rm EM}=1$.
Here, $\mathcal L_{\rm Adam}$ is the reconstruction residual and
$\mathcal L_{\rm EM}$ is already an averaged negative log-likelihood. The
candidate is accepted when $d_a\leq\eta_a$ and rejected otherwise. Each
$\eta_a$ is calibrated to balance acceptance of exact or partially correct nominal candidates
against random-candidate rejection.

The fitting-based attack exploits the available structural leakage but retains
two limitations. First, Adam and \gls{em} still solve a non-convex
joint key--\gls{an} problem and may converge to equivalent waveform
decompositions. Second, the \emph{shared key} has only $k$ unknown phases, whereas the
\emph{smooth key} has $3k$ generator parameters and the \emph{seed-expanded key} provides no
differentiable phase relation. These fitting algorithms cannot directly recover
the \emph{seed-expanded phase key}. Once the nominal root is determined, Eve must search
the $\lambda$-bit key seed; structural leakage may reject candidates early but
does not order the pseudorandom space. This unstructured search costs
$\Omega(2^\lambda)$ classical tests in the worst case (about $2^{\lambda-1}$
on average)\textcolor{black}{, or $\Theta(2^{\lambda/2})$ Grover queries}.

\subsection{Symbol Matching and Sequence Localization}


\label{subsec:symbol_matching_localization}
After recovering multiple imperfect copies of the same sensing sequence, Eve
may localize the sequence by maximizing their symbol-agreement score. Let
$A\in\{0,\ldots,M-1\}$ be a transmitted symbol and let $\hat A_1$ and
$\hat A_2$ be its estimates from two independently recovered copies. We
assume a symmetric $M$-ary error model with \gls{ser} $p_s$,
\begin{equation*}
\Pr(\hat A_j=b\mid A=a)
=
\begin{cases}
1-p_s, & b=a,\\
p_s/(M-1), & b\neq a,
\end{cases}
\qquad j\in\{1,2\}.
\end{equation*}
When aligned, estimates agree if both are correct or
select the same wrong symbol. Conditional independence gives
\begin{equation}
\begin{aligned}
p_{\rm a}(p_s)
&=\Pr(\hat A_1=\hat A_2\mid A=a) \\
&=\sum_{b=0}^{M-1}
\Pr(\hat A_1=b\mid A=a)
\Pr(\hat A_2=b\mid A=a) \\
&=(1-p_s)^2
+(M-1)\left(\frac{p_s}{M-1}\right)^2 \\
&=(1-p_s)^2+\frac{p_s^2}{M-1}.
\end{aligned}
\label{eq:agreement_from_ser}
\end{equation}
\textcolor{black}{The mixed cases, in which one estimate is correct and the other is wrong,
cannot produce agreement because an erroneous estimate excludes the true
symbol.} At an unrelated alignment, the underlying symbols are independent;
for a uniform alphabet and the symmetric error model, the recovered symbols
are also marginally uniform, and hence agree with probability $1/M$.
For a length-$N$ sequence, the correct-alignment score is approximately
$S_{K_0}\sim\operatorname{Binomial}(N,p_{\rm a})$, whereas an unrelated
symbol offset has $S_K\sim\operatorname{Binomial}(N,1/M)$. With
$\hat K=\arg\max_K S_K$ and true symbol offset $K_0$,
the localization {\color{black}\gls{mae}} is
\begin{equation}
D_{\rm MAE}(p_s)
=\mathbb{E}\{|\hat K-K_0|\}
=\sum_K |K-K_0|\Pr(\hat K=K).
\label{eq:ser_to_location_mae}
\end{equation}
The mapping is not determined by \gls{ser} alone, since it also depends on
$N$, $M$, and the search interval.

\textbf{Coherent sensing attack with phase-key mismatch and residual \glsentryshort{an}:}
We relate key-recovery accuracy to correct-delay locking through a power-domain
approximation. For echo $b$, let the residual key error be
$\delta_{\phi,b}\sim\mathcal{N}(0,\sigma_\phi^2)$. Its coherent amplitude and
power factors are
\begin{equation}
\rho_\phi=\mathbb{E}\!\left[e^{j\delta_{\phi,b}}\right]
=e^{-\sigma_\phi^2/2},\quad
G_\phi=|\rho_\phi|^2=e^{-\sigma_\phi^2}.
\end{equation}
{\color{black}For $B$ independently accumulated echoes with per-echo linear \gls{snr}
$\gamma$, the normalized thermal-noise, key-mismatch, and residual-\gls{an}
powers are approximated as
\begin{align*}
P_{\rm th}&=\frac{1}{(1-\beta)B\gamma},\enspace
P_{\rm key}&=\frac{1-G_\phi}{B},\enspace
P_{{\rm AN}}&=\frac{\beta(1-G_\phi)}{(1-\beta)B}.
\end{align*}
}
\begin{equation}
\gamma_{\rm eff}^{({\rm key+AN})}
=\frac{G_\phi}{P_{\rm th}+P_{\rm key}+P_{{\rm AN}}}.
\label{eq:key_error_noncentrality}
\end{equation}
This expression accounts for phase-key mismatch and the resulting imperfect
tangent-\gls{an} cancellation.
The correct-lock probability is evaluated against the ideal-reference detector
under the same channel, front end, delay search, and threshold:
\begin{equation}
P_{\rm corr}(\sigma_\phi)
\approx\mathcal{P}_{\rm corr}^{(0)}\!\left(\gamma_{\rm eff}\right),
\label{eq:key_error_pcorr_model}
\end{equation}
where $\mathcal{P}_{\rm corr}^{(0)}(\cdot)$ denotes the channel-aware baseline
correct-lock curve. This approximation assumes independent zero-mean key
errors and echo realizations.

\begin{figure*}[!t]
    \centering
    \subfloat[]{\includegraphics[width=0.232750\textwidth]{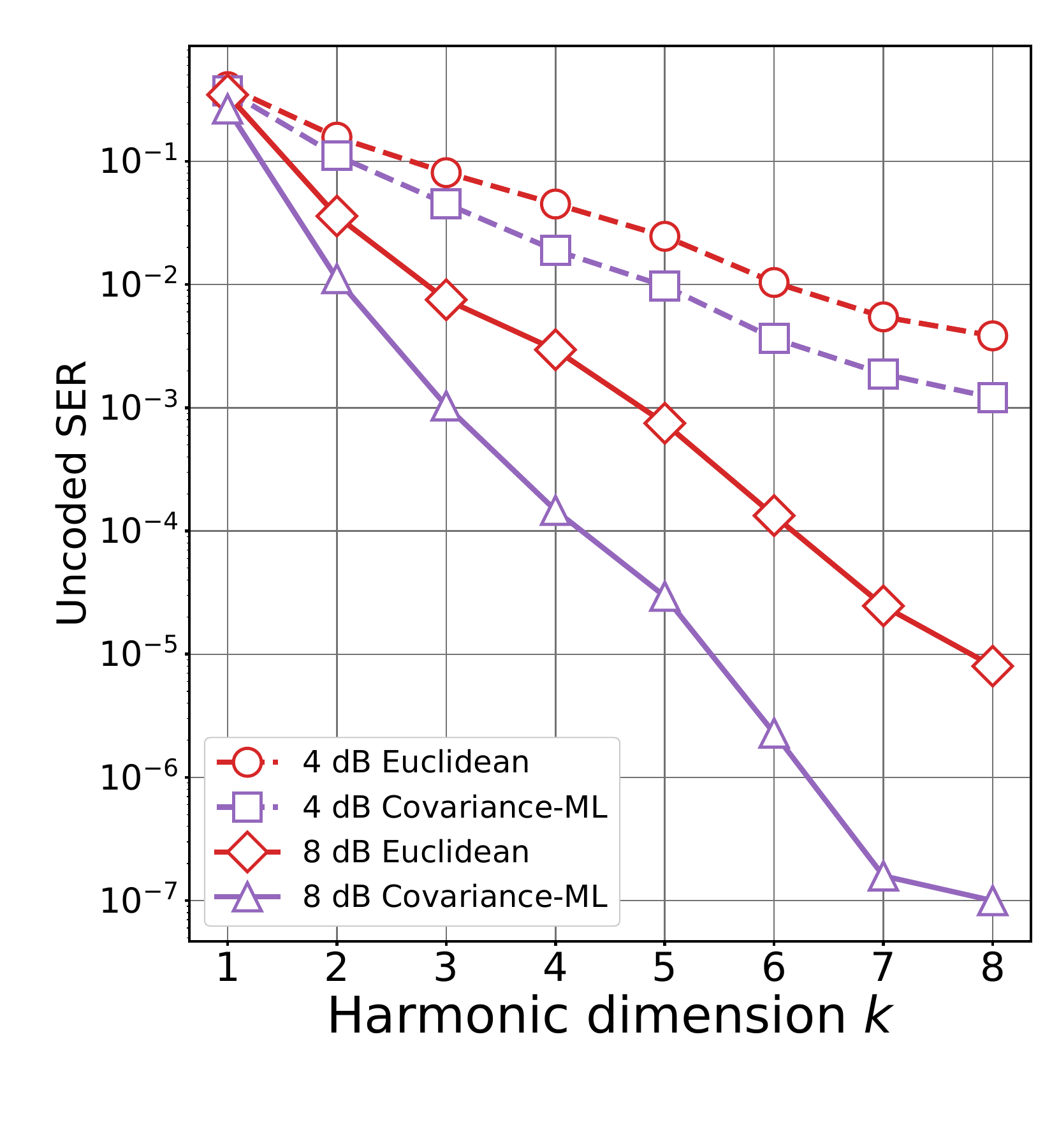}}
    \hfill
    \subfloat[]{\includegraphics[width=0.232750\textwidth]{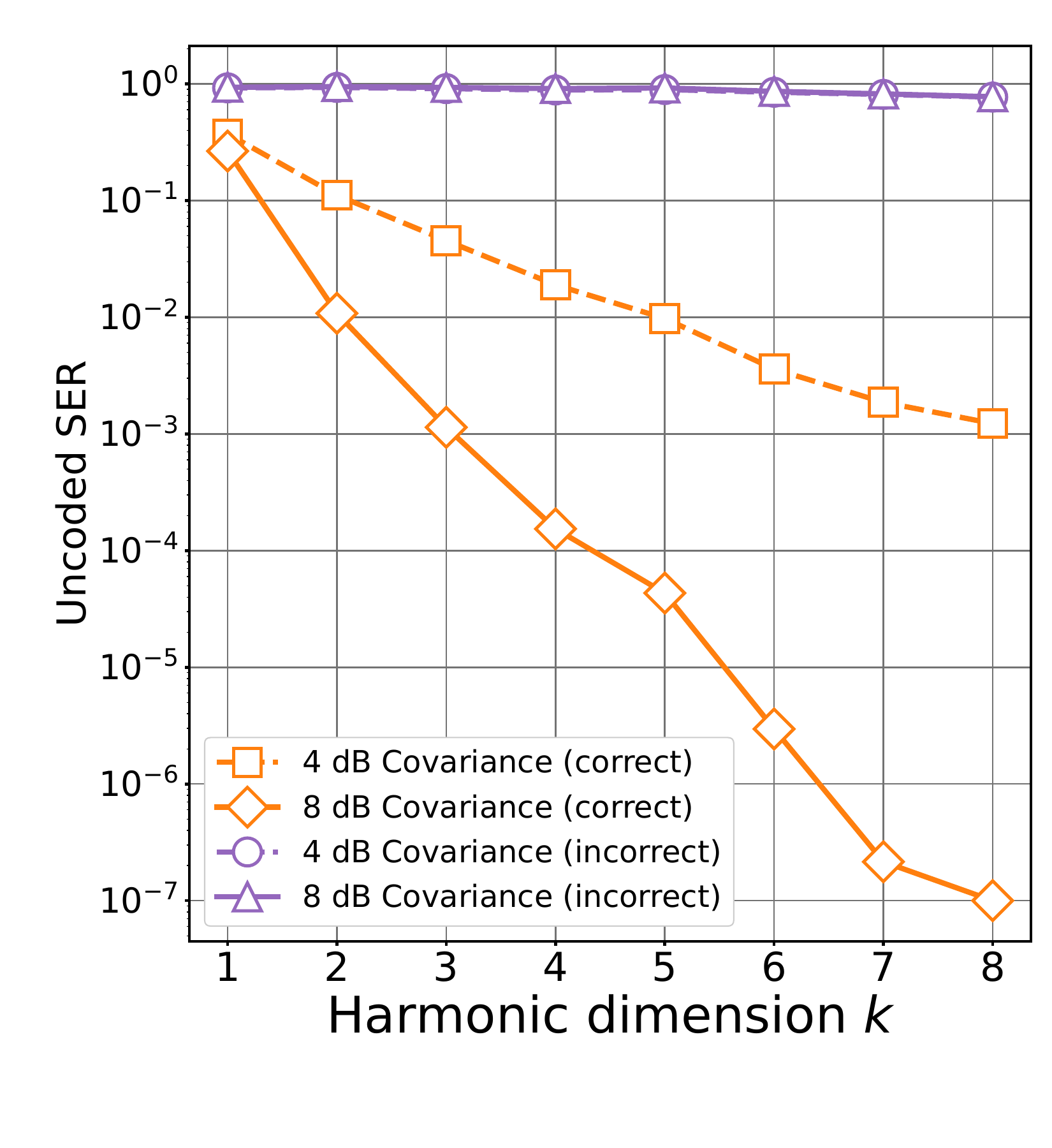}}
    \hfill
    \subfloat[]{\includegraphics[width=0.232750\textwidth]{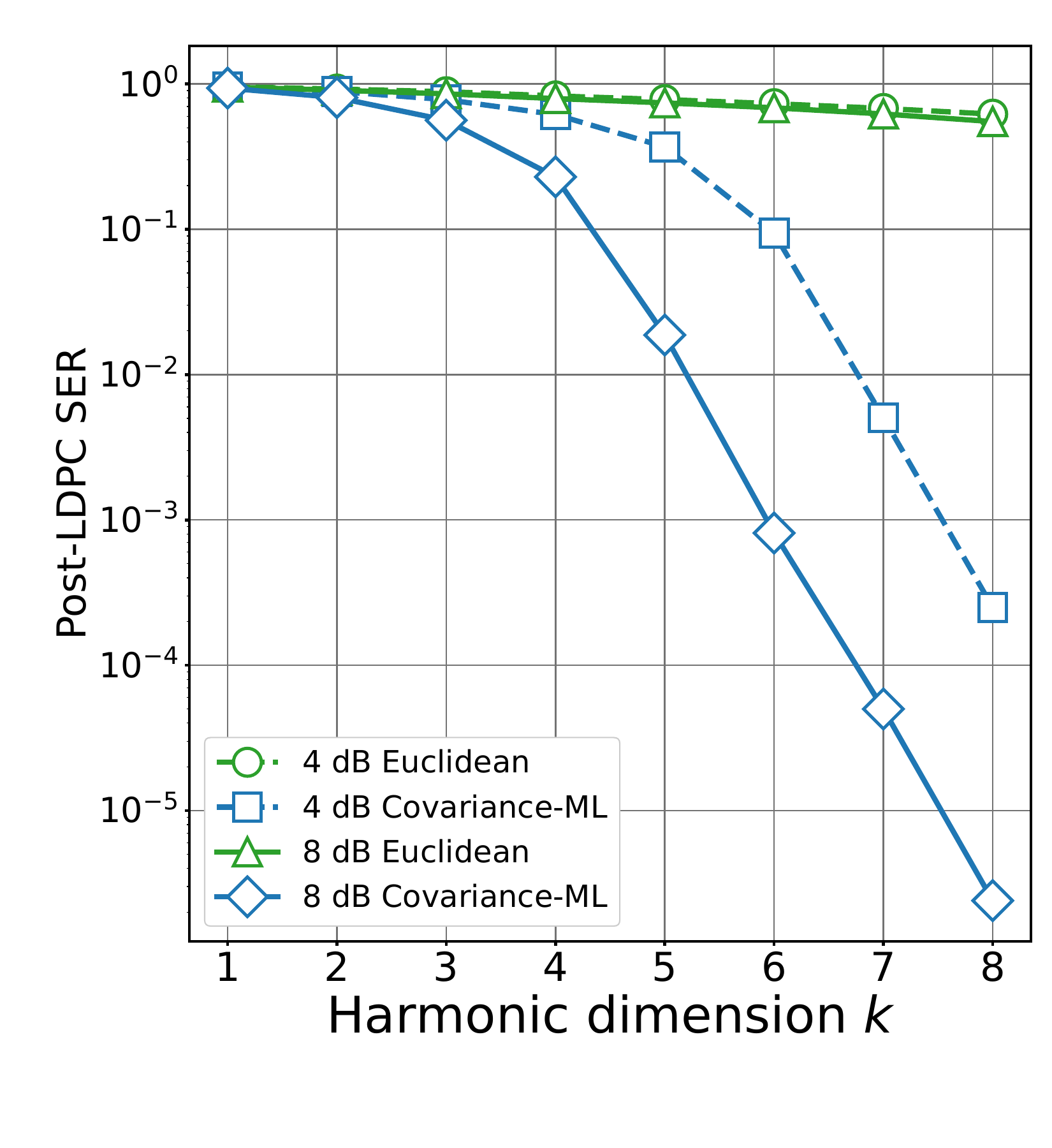}}
    \hfill
    \subfloat[]{\includegraphics[width=0.232750\textwidth]{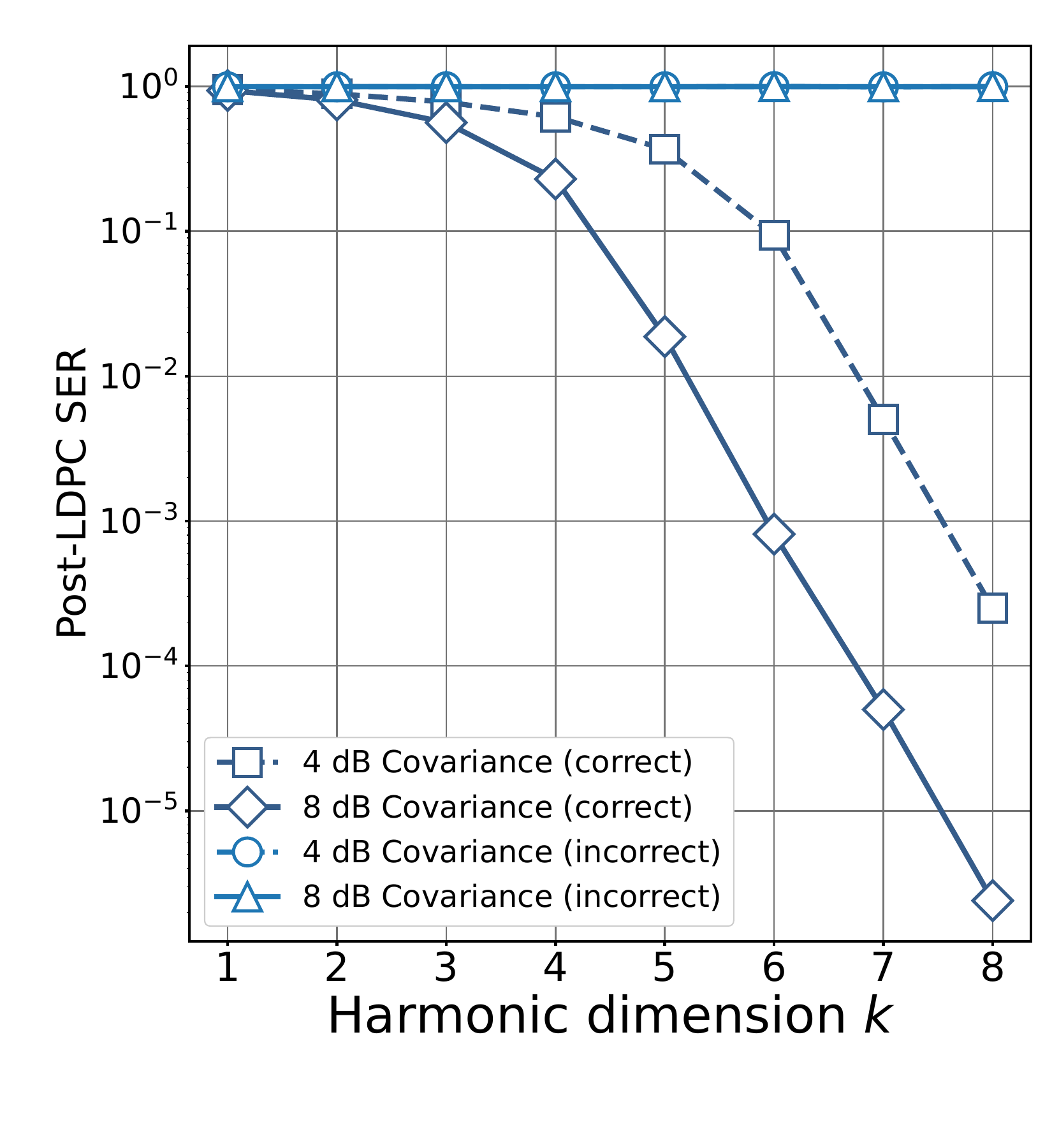}}
    \caption{\glsentryshort{ser} versus harmonic dimension at $4$ and $8$ dB.
    (a)--(b) Uncoded \glsentryshort{qpsk}: unkeyed Euclidean and
    covariance-aware demodulation, and keyed covariance-aware demodulation
    with and without the phase key. The $M=4$ phases repeat cyclically across
    the $k$ harmonic components. (c)--(d) Rate-$1/2$ \glsentryshort{ldpc}-coded 64-PSK under the
corresponding unkeyed and keyed settings.}
\label{fig:communication_ber_combined}
\end{figure*}
\section{Numerical Results}
We evaluate security under increasing Eve knowledge and capability.
After validating communication and waveform design, we test blind
cross-burst correlation. With waveform and key-generation rules known, Eve
uses an offline \gls{cnn} to identify nominal roots; with additional nominal-symbol
knowledge, she fits the phase key online using Adam or phase-aware \gls{em}.
Finally, we evaluate the coherent sensing attack through detection and
delay locking with imperfect keys.

\subsection{Validation of Secure Communication and Waveform Design}

We evaluate communication security using \gls{qpsk} for sensing and control
and higher-order modulation for payload. A naive Eve applies Euclidean
demapping, whereas an informed Eve uses covariance-aware demapping but lacks
the phase key.
\textcolor{black}{Tangent-\gls{an} and phase keys are independently regenerated,
with $\beta=0.5$ as in~\cite{han2026matched}.}

Fig.~\ref{fig:communication_ber_combined}(a)--(b) shows the uncoded
\gls{qpsk} results. Increasing $k$ lowers the \gls{ser} by adding harmonic
features and redundancy, while covariance-aware demapping outperforms the
Euclidean metric. With keying, the legitimate receiver retains the correct
codebook, whereas Eve's mismatched codebook keeps her \gls{ser} high.
\textcolor{black}{At
$k=5$ and $8$ dB, Eve's \gls{ser} is $0.7498$, $0.7497$, and $0.7484$ for
the \emph{shared}, \emph{smooth}, and \emph{seed-expanded} key-generation rules, respectively.}
Their communication performance is therefore similar; for simplicity,
Fig.~\ref{fig:communication_ber_combined} uses the \emph{shared key}.
Fig.~\ref{fig:communication_ber_combined}(c)--(d) shows the corresponding
rate-$1/2$ \gls{ldpc}-coded 64-PSK results. The same trend remains, although
the denser constellation increases sensitivity to noise and tangent
\gls{an}. Larger $k$ therefore improves reliability at the cost of additional
signal dimensions\textcolor{black}{, a suitable tradeoff for short sensing and control signals}.
\textcolor{black}{Overall, key mismatch maintains a high error rate at Eve even when she knows
the receiver architecture and applies covariance-aware demapping.}

\subsubsection{Waveform Design and Legitimate Sensing}
We validate sensing-waveform design for cooperating \glspl{bs}.
Table~\ref{tab:opt_vs_random_corr} shows that optimization raises \gls{psr}
above the random baseline, reaching around $20$~dB for most $k$.
The dips at $k=4$ and $k=8$ are a
\textcolor{black}{structural effect of the \gls{qpsk} alphabet: for $\phi_n = i\pi/2$, the harmonic
sum $\sum_{m=1}^{k} e^{jm\phi_n}$ has period $4/\gcd(i,4)$ in $m$ and
vanishes when $k$ is divisible by this period, leaving only
$\phi_n=0$ non-degenerate and reducing diversity.}

\begin{table}[!t]
    \centering
    \caption{\textcolor{black}{\glsentryshort{psr} (dB) of optimized and random
    \glsentryshort{qpsk} waveforms: $N=64$, \glsentryshort{snr} $=5$ dB,
    $\beta=0.5$, 2000 \glsentryshort{mc} runs.}}
    \label{tab:opt_vs_random_corr}
    \tabletextsize
    \setlength{\tabcolsep}{1.8pt}
    \renewcommand{\arraystretch}{1.05}
    \begin{tabular}{lcccccccc}
        \toprule[1.2pt]
        \rowcolor{gray!20}
        \textbf{$k$} & \textbf{1} & \textbf{2} & \textbf{3} & \textbf{4} & \textbf{5} & \textbf{6} & \textbf{7} & \textbf{8} \\
        \midrule
        \rowcolor{gray!8}
        \textbf{Opt. \glsentryshort{psr}} & 20.95 & 21.86 & 19.65 & 17.40 & 20.52 & 21.47 & 19.11 & 17.36 \\
        \rowcolor{gray!15}
        \textbf{Rand. \glsentryshort{psr}} & 16.29 & 13.90 & 13.51 & 7.67 & 12.29 & 10.80 & 12.16 & 8.67 \\
        \bottomrule[1.2pt]
    \end{tabular}
\end{table}
\subsubsection{Repeated-Burst Waveform Protection}
Here, Eve observes repeated received waveforms but does not know their
generative structure and cannot reliably decode their symbols, as demonstrated
above. Her detector therefore cross-correlates the observations.
The same nominal sequence is
transmitted with independently refreshed phase keys and \gls{an} realizations.
Table~\ref{tab:keyed_burst_beta} compares optimized and random sequences using
the detection probability $P_D$ under the peak-threshold rule in
\eqref{eq:peak_search}. The unprotected optimized sequence
(\textbf{Opt. No}) gives $P_D=1.0$, versus $0.05$ for \textbf{Rand. No}
(different random-sequence realizations).
Reusing the same key with independent \gls{an} at $\beta=0.5$
(\textbf{Opt. A+SK}) gives $P_D=0.925$. Refreshing the keys alone reduces
$P_D$ to $0.386$; adding \gls{an} lowers it below the
unprotected random baseline at $\beta=0.3$, with no detections observed for
\textcolor{black}{$\beta\geq0.6$, as in the protected random baseline
(\textbf{Rand. A+K}, $\beta=0.5$), highlighting the importance of key
refreshing.} Although the protected optimized sequence retains a higher
\gls{psr}, this ratio describes correlation shape: detection also requires
sufficient peak magnitude. Independent keys suppress cross-burst alignment
and the resulting peak, reducing detection despite the higher \gls{psr}.

\begin{table}[!t]
    \centering
    \caption{\textcolor{black}{Repeated-burst keyed \glsentryshort{qpsk}
    cross-correlation: $N=64$, $k=5$, \glsentryshort{snr} $=5$ dB,
    $P_{\rm FA}=0.05$.}}
    \label{tab:keyed_burst_beta}
    \tabletextsize
    \setlength{\tabcolsep}{4.1pt}
    \renewcommand{\arraystretch}{1}
    \begin{tabular*}{0.84\columnwidth}{@{\extracolsep{\fill}}C{0.12\columnwidth}ccc|C{0.12\columnwidth}ccc}
        \toprule[1.2pt]
        \rowcolor{gray!20}
        \textbf{$\beta$} & \textbf{Peak} & \textbf{\glsentryshort{psr}} & \textbf{$P_D$}
        & \textbf{$\beta$} & \textbf{Peak} & \textbf{\glsentryshort{psr}} & \textbf{$P_D$} \\
        \midrule
        \rowcolor{gray!5}
        0.0 & 0.262 & 5.620 & 0.386 & 0.5 & 0.143 & 2.350 & 0.002 \\
        \rowcolor{gray!5}
        0.1 & 0.235 & 5.280 & 0.172 & 0.6 & 0.124 & 1.480 & 0.000 \\
        \rowcolor{gray!5}
        0.2 & 0.211 & 4.890 & 0.074 & 0.7 & 0.116 & 1.000 & 0.000 \\
        \rowcolor{gray!5}
        0.3 & 0.187 & 4.160 & 0.016 & 0.8 & 0.112 & 0.810 & 0.000 \\
        \rowcolor{gray!5}
        0.4 & 0.160 & 3.070 & 0.002 & 0.9 & 0.112 & 0.690 & 0.000 \\
        \midrule
        \rowcolor{gray!15}
        $\textbf{Rand. No}$& 0.202 & 0.780 & 0.050 & $\textbf{Rand. A+K}$& 0.105 & 0.626 & 0.000 \\
        \rowcolor{gray!15}
        \midrule
        \rowcolor{gray!8}
        $\textbf{Opt. No}$& 0.725 & 6.880 & 1.000 & $\textbf{Opt. A+SK}$& 0.369 & 6.976 & 0.925\\
        \bottomrule[1.2pt]
    \end{tabular*}
\end{table}

\subsection{Validation Against Eve's Inference Attacks}
\subsubsection{Phase-Key Mismatch and Search Complexity}
We first evaluate an informed Eve that reconstructs the phase key with a
residual average error $\sigma_\phi${\color{black}, using the covariance-aware receiver
with $\beta=0.5$ and \gls{snr} $=1$ dB}. Table~\ref{tab:key_mismatch_ser_k5}
shows that the resulting \gls{ser} increases from $0.055$ with the exact key
to $0.581$ at $\sigma_\phi=0.5\pi$. If Eve knows only the phase-key space,
{\color{black}each of the $kN$ phases is quantized into
$Q=\lceil2/\sigma_\phi\rceil$ values. Scoring every candidate array by
demodulating all $N$ \gls{qpsk} symbols, with four metrics per symbol, gives
\begin{equation}
C_{\rm Eve}=4NQ^{kN}
=4N\left\lceil\frac{2}{\sigma_\phi}\right\rceil^{kN}
\label{eq:eve_grid_complexity}
\end{equation}
metric evaluations per burst. This count applies without knowledge of the
key-generation rule.} {\color{black}Table~\ref{tab:key_mismatch_ser_k5} also reports this
blind-search cost and the localization \gls{mae} from \eqref{eq:ser_to_location_mae},
using $200{,}000$ \gls{mc} runs. The 64-symbol sensing sequence is
embedded between 64 random symbols on each side. Eve slides a 64-symbol
reference across the 192-symbol stream and selects the best-matching
start position.}
\begin{table}[!t]
    \centering
    \caption{\textcolor{black}{\glsentryshort{ser}, blind-search cost, and localization \gls{mae} under phase-key mismatch:
    \textcolor{black}{$N=64$, $k=5$, \glsentryshort{snr} $=1$ dB, $\beta=0.5$}.}}
    \label{tab:key_mismatch_ser_k5}
    \tabletextsize
    \setlength{\tabcolsep}{1.0pt}
    \renewcommand{\arraystretch}{1.05}
    \begin{tabular}{@{}>{\centering\arraybackslash}p{0.17\columnwidth}*{6}{>{\centering\arraybackslash}p{0.125\columnwidth}}@{}}
        \toprule
        \rowcolor{gray!20}
        \textbf{$\sigma_\phi$ ($\pi$)} &
        \textbf{0} & \textbf{0.05} & \textbf{0.10} & \textbf{0.15} &
        \textbf{0.20} & \textbf{0.25} \\
        \midrule
        \rowcolor{gray!8}
        \textbf{\glsentryshort{ser}} &
        0.055 & 0.060 & 0.074 & 0.112 & 0.152 & 0.238 \\
        \rowcolor{gray!12}
        \textbf{$D_{\rm MAE}$} &
        0.00 & 0.00 & 0.00 & 0.00 & 0.00 & 0.05 \\
        \rowcolor{gray!15}
        \textbf{$\log_{10}C_{\rm Eve}$} &
        -- & 515.1 & 418.7 & 369.2 & 322.4 & 291.4 \\
        \bottomrule
    \end{tabular}
    \vspace{2pt}

    \begin{tabular}{@{}>{\centering\arraybackslash}p{0.17\columnwidth}*{6}{>{\centering\arraybackslash}p{0.125\columnwidth}}@{}}
        \toprule
        \rowcolor{gray!20}
        \textbf{$\sigma_\phi$ ($\pi$)} &
        \textbf{0.30} & \textbf{0.35} & \textbf{0.40} &
        \textbf{0.45} & \textbf{0.50} & \textbf{--} \\
        \midrule
        \rowcolor{gray!8}
        \textbf{\glsentryshort{ser}} &
        0.289 & 0.377 & 0.529 & 0.527 & 0.581 & -- \\
        \rowcolor{gray!12}
        \textbf{$D_{\rm MAE}$} &
        0.74 & 9.27 & 29.37 & 29.22 & 31.22 & -- \\
        \rowcolor{gray!15}
        \textbf{$\log_{10}C_{\rm Eve}$} &
        272.8 & 251.4 & 226.1 & 226.1 & 195.1 & -- \\
        \bottomrule
    \end{tabular}
\end{table}

\begin{figure*}[!t]
\centering
\includegraphics[width=0.451250\textwidth]{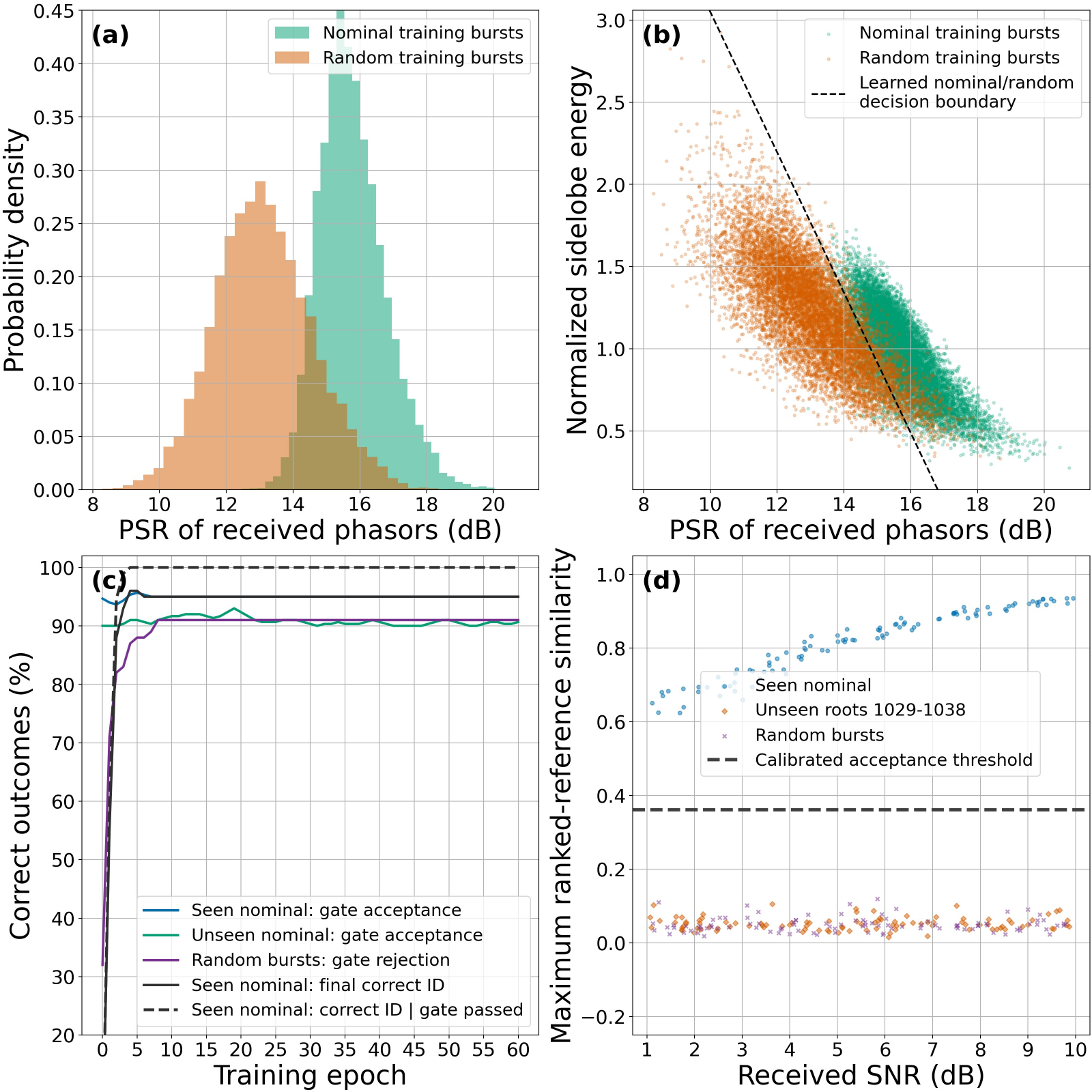}
\hspace{0.025\textwidth}
\includegraphics[width=0.451250\textwidth]{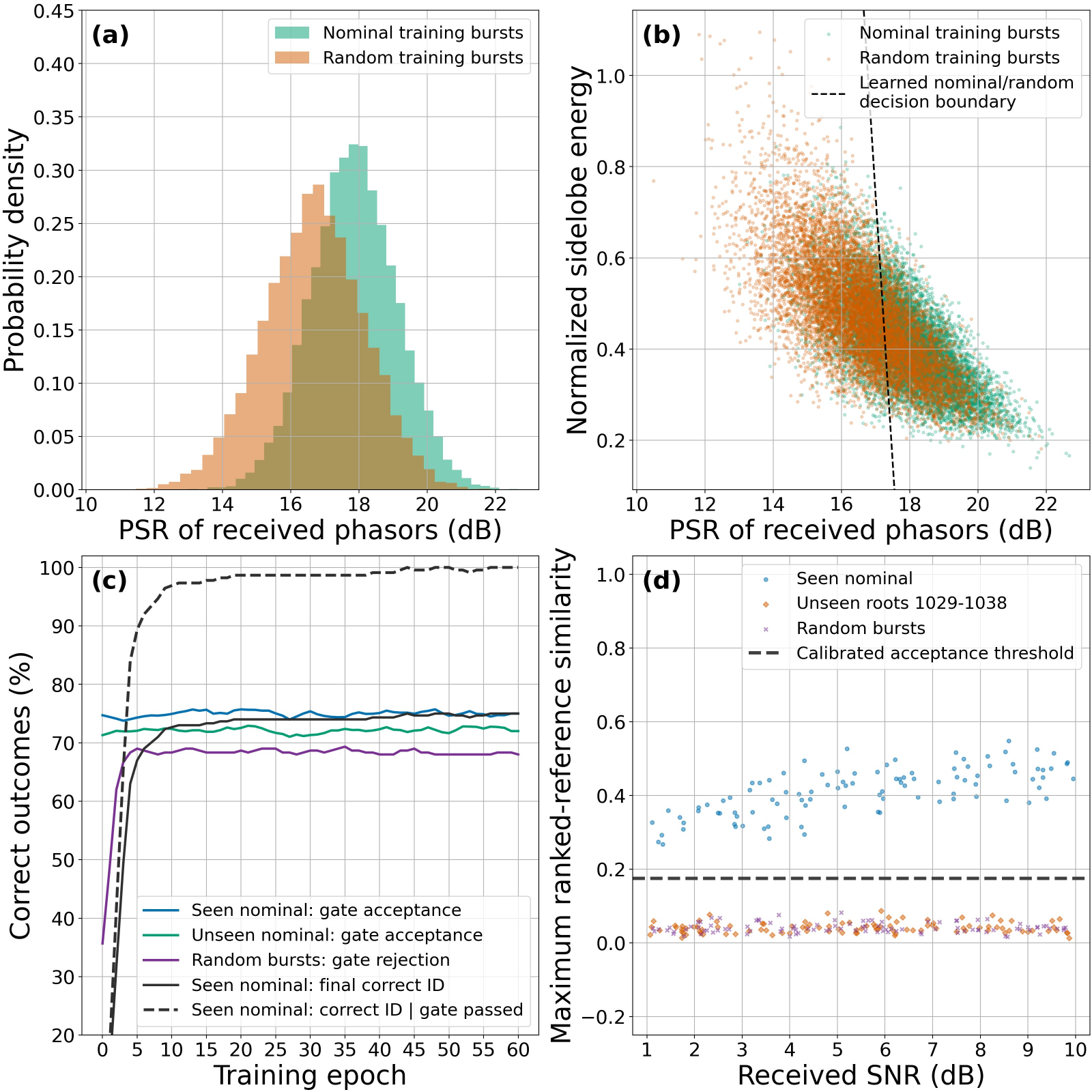}\\[4mm]
\includegraphics[width=0.451250\textwidth]{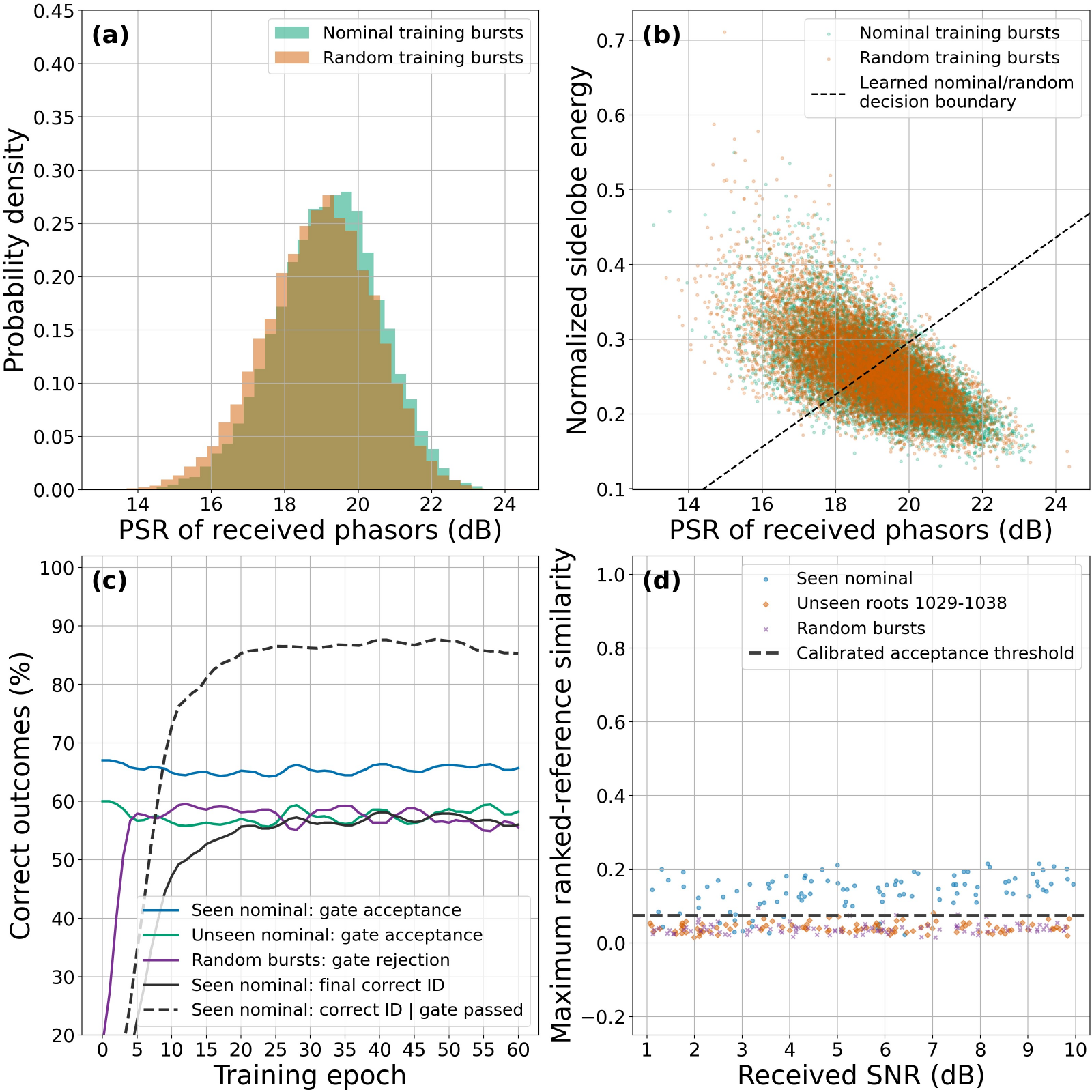}
\hspace{0.025\textwidth}
\includegraphics[width=0.451250\textwidth]{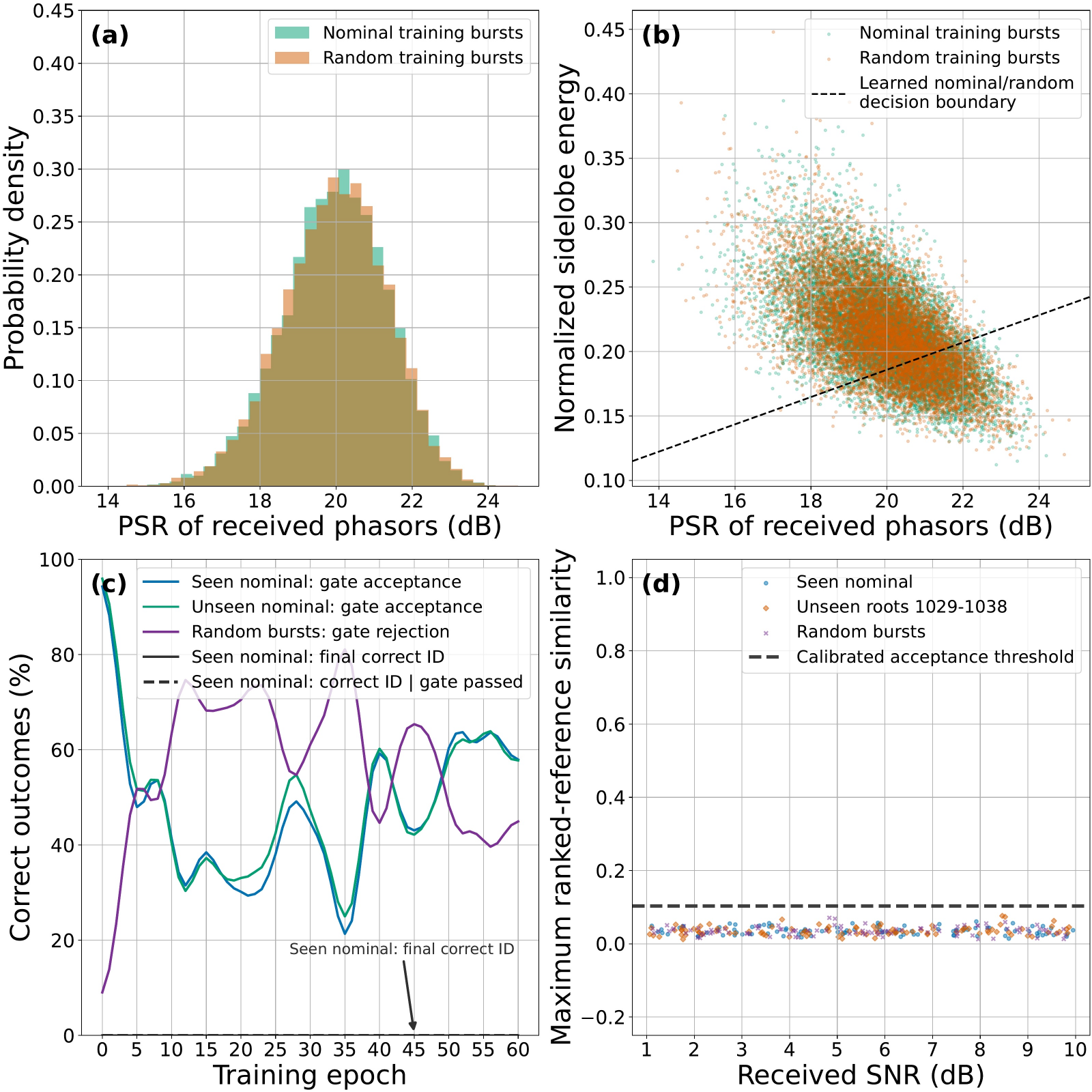}
\caption{Offline sequence detection and root identification. Top left:
unprotected baseline ($\beta=0$). Top right: \emph{burst-shared lookup-table key}.
Bottom left: \emph{smooth position-dependent key}. Bottom right: \emph{seed-expanded
pseudorandom key} \textcolor{black}{with a 1024-seed space}. Within each block, (a) shows
empirical \glsentryshort{psr} distributions, (b) the learned gate boundary,
(c) gate outcomes and root-identification accuracy, and (d) ranked-reference
similarity with the calibrated acceptance threshold.}
\label{fig:eve_detector_baselines}
\label{fig:eve_detector_keyed}
\label{fig:eve_detector_csprng}
\end{figure*}

\begin{figure*}[!t]
    \centering
    \subfloat[]{\includegraphics[width=0.232750\textwidth]{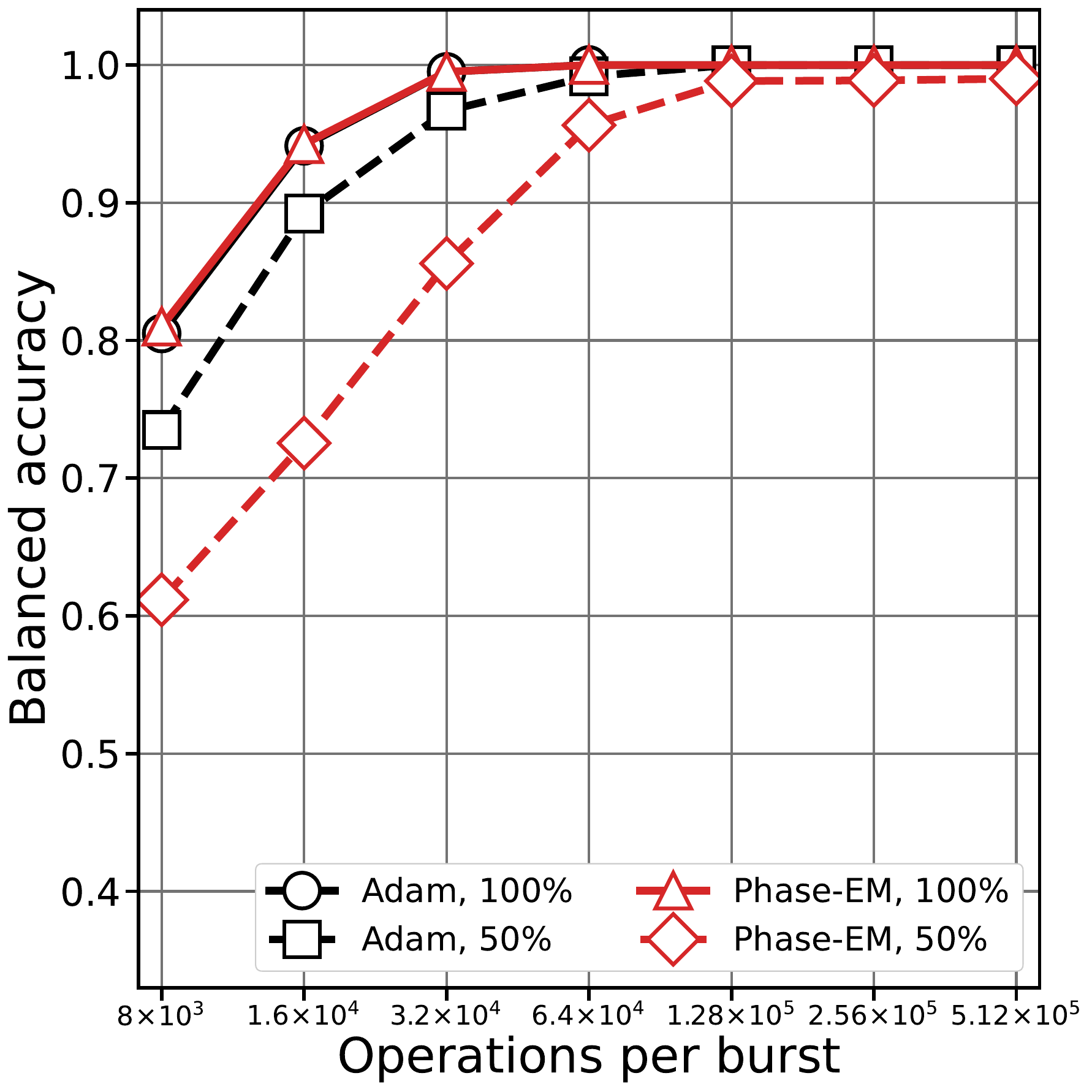}}
    \hfill
    \subfloat[]{\includegraphics[width=0.232750\textwidth]{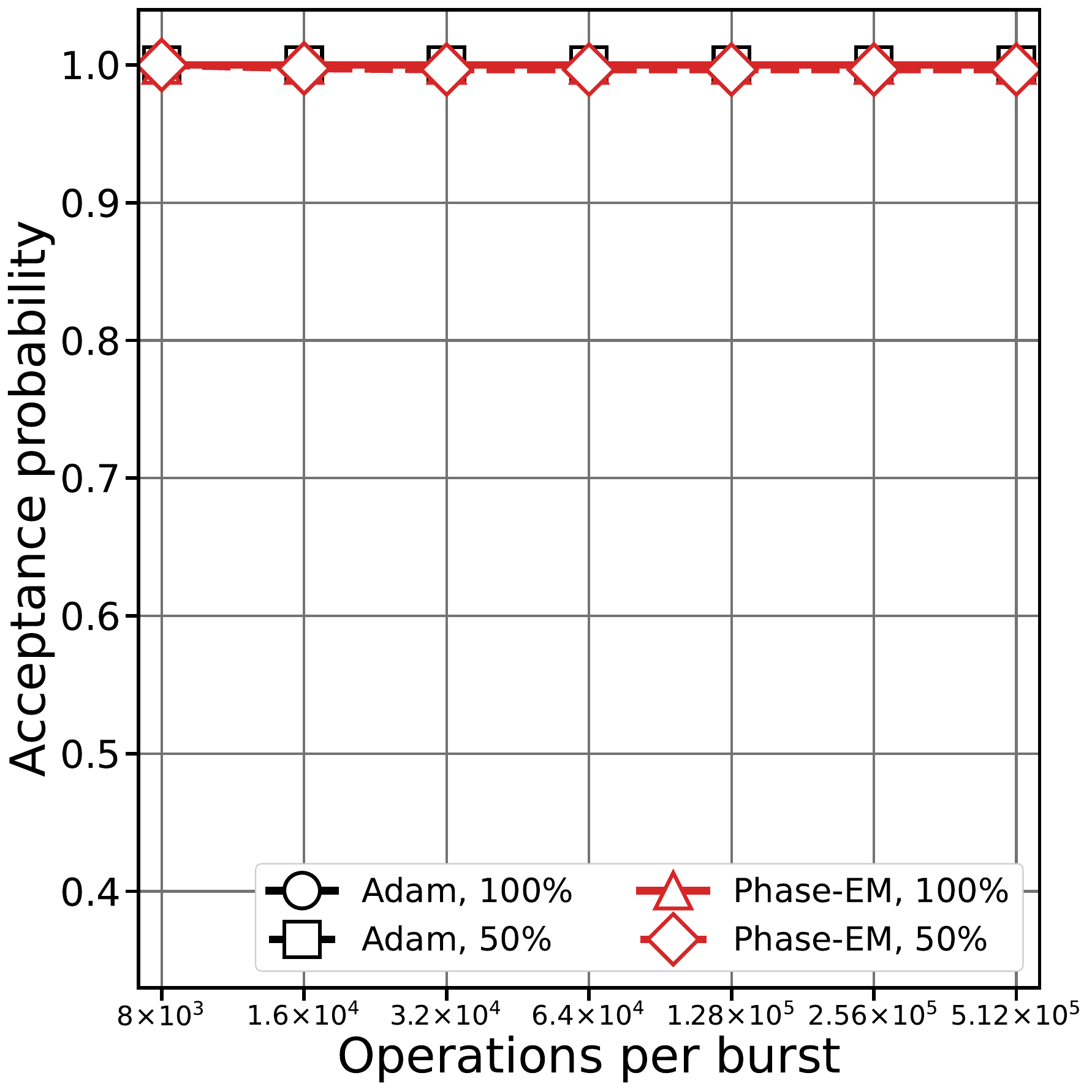}}
    \hfill
    \subfloat[]{\includegraphics[width=0.232750\textwidth]{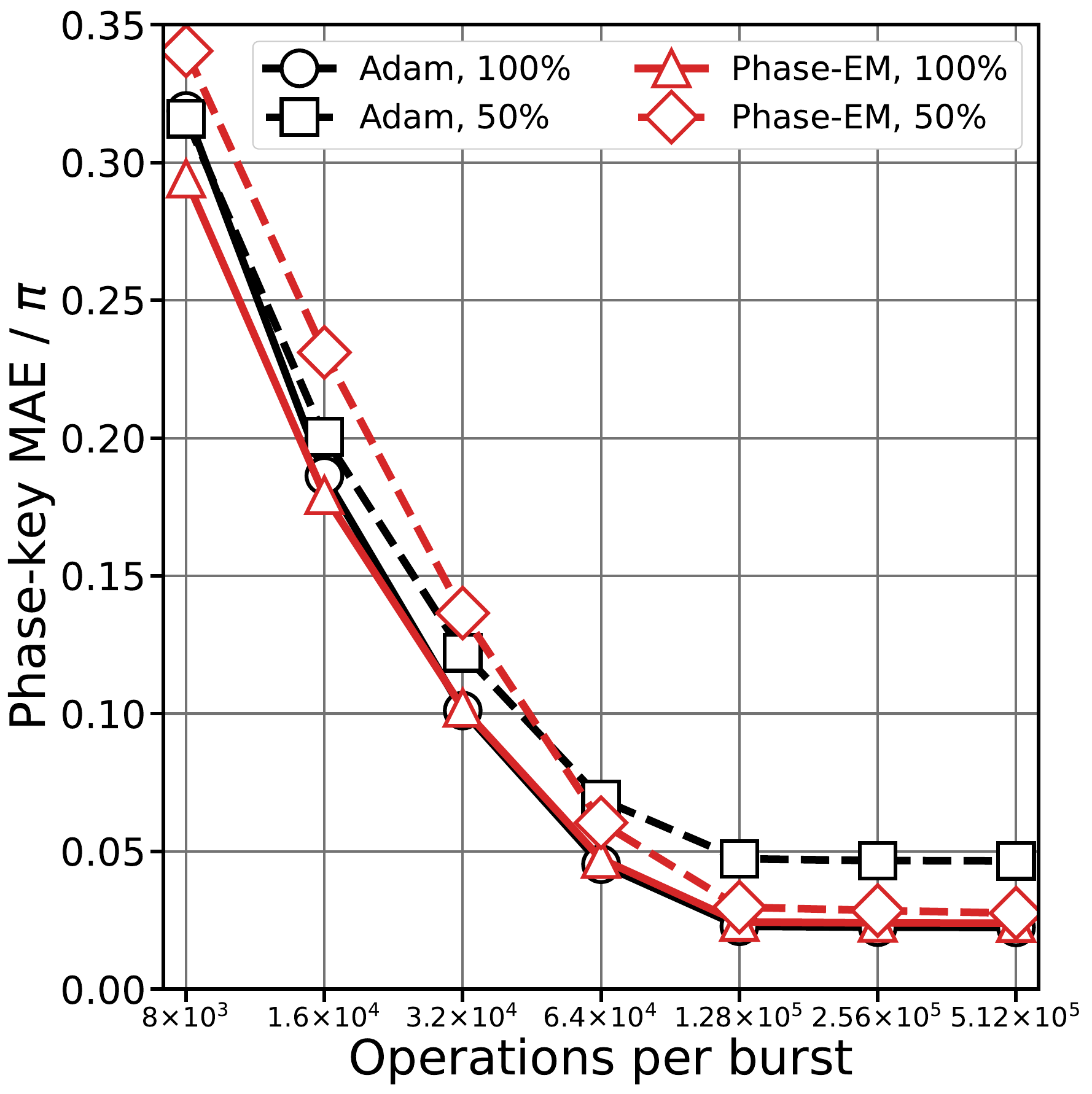}}
    \hfill
    \subfloat[]{\includegraphics[width=0.232750\textwidth]{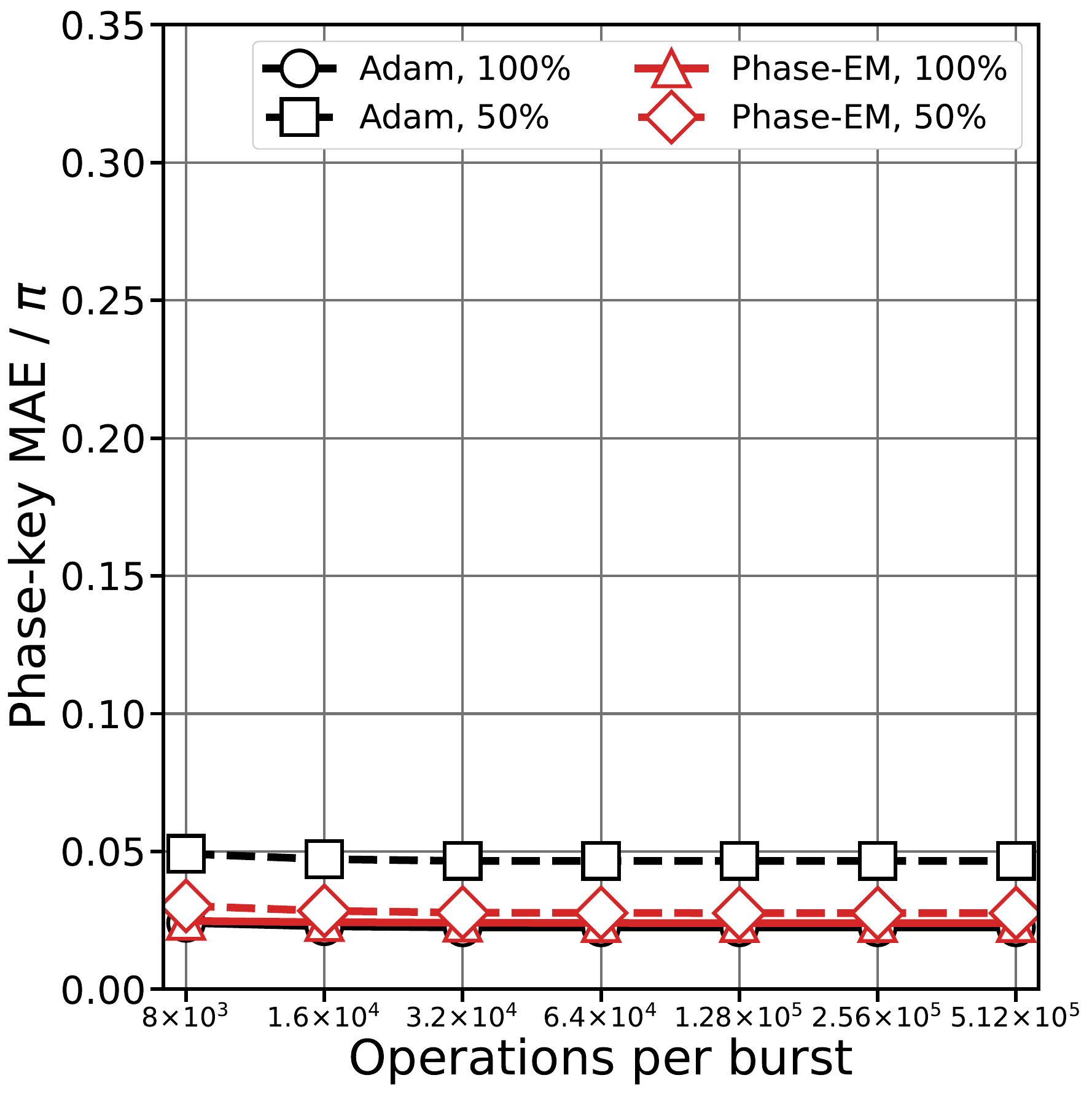}}
    \caption{Online fitting-based detection under the \emph{shared key} versus the
    operation budget per burst: balanced detection accuracy (a) without
    harmonic-structure initialization and nominal-sequence acceptance (b) with
    it; phase-key \gls{mae} (c) without and
    (d) with harmonic-structure initialization.}
    \label{fig:fitting_detector_validation}
\end{figure*}

\begin{figure*}[!t]
    \centering
    \subfloat[]{\includegraphics[width=0.232750\textwidth]{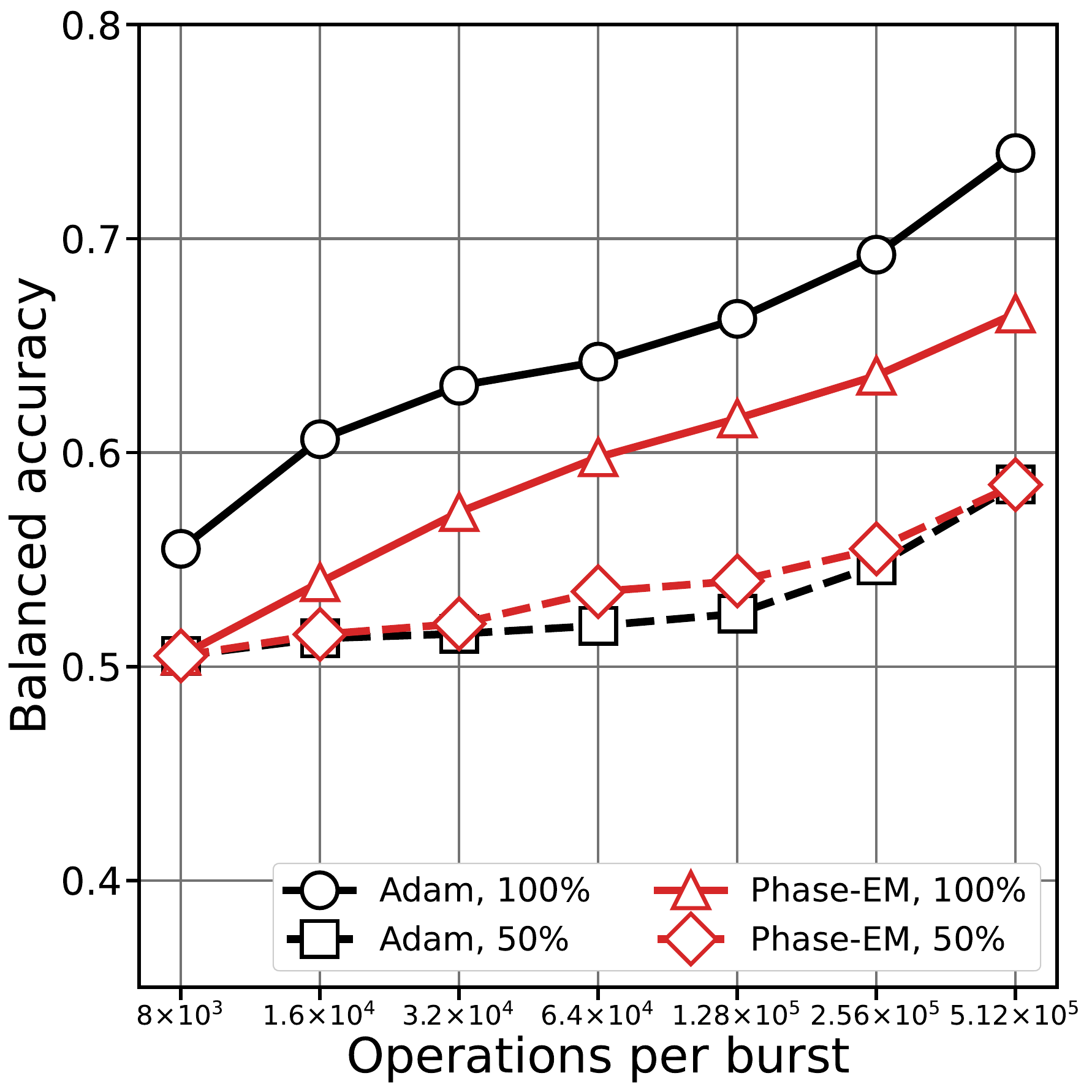}}
    \hfill
    \subfloat[]{\includegraphics[width=0.232750\textwidth]{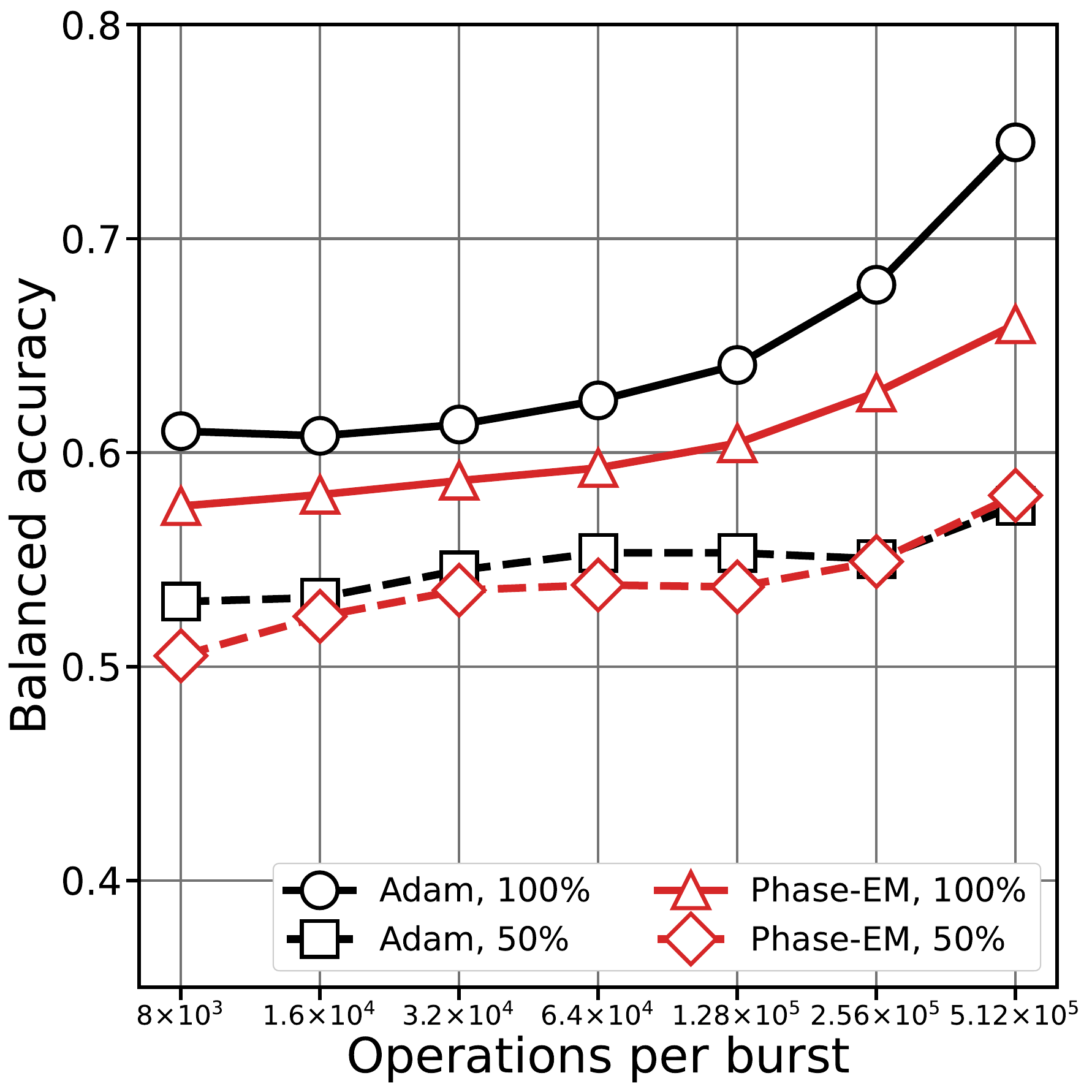}}
    \hfill
    \subfloat[]{\includegraphics[width=0.232750\textwidth]{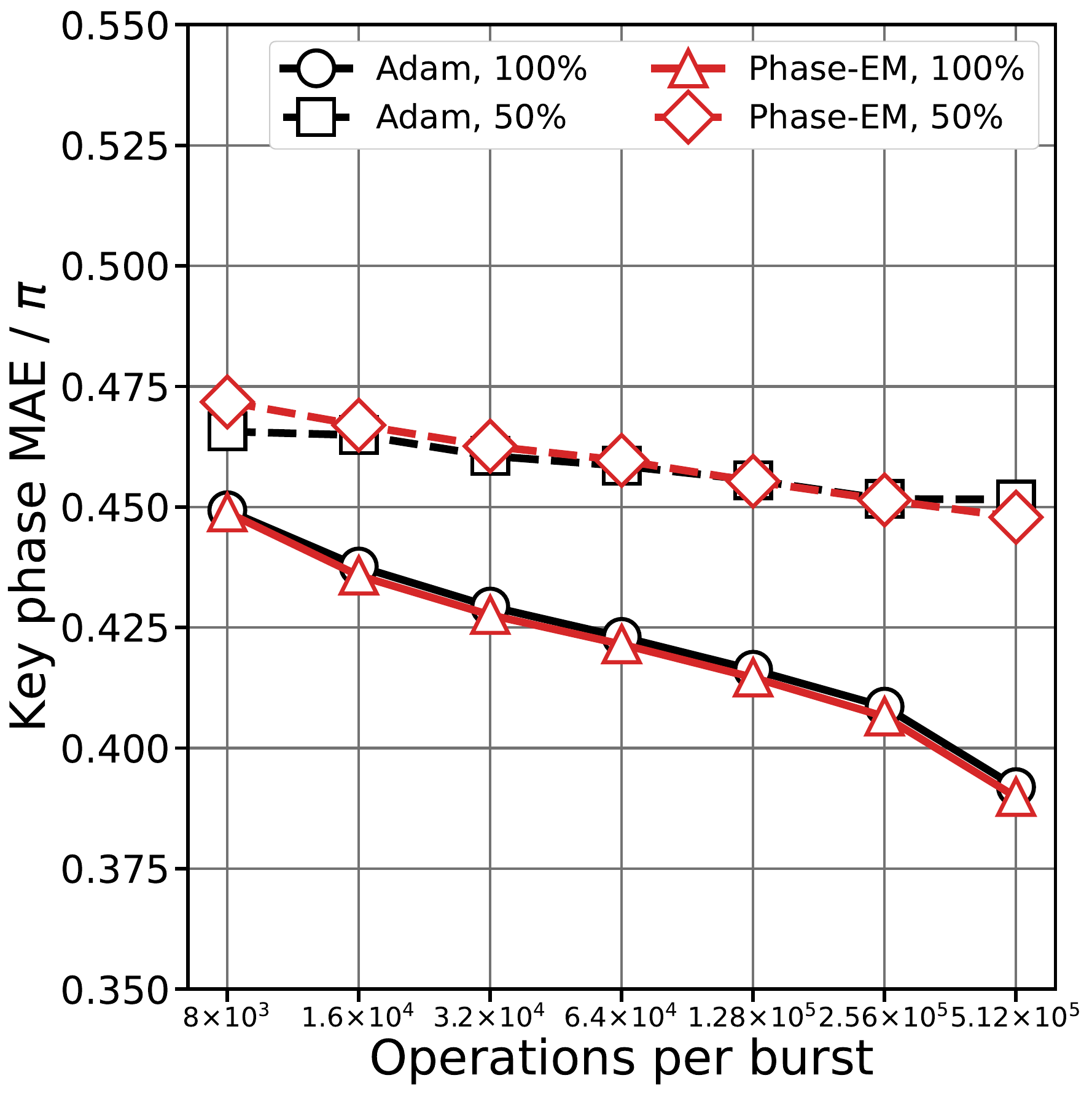}}
    \hfill
    \subfloat[]{\includegraphics[width=0.232750\textwidth]{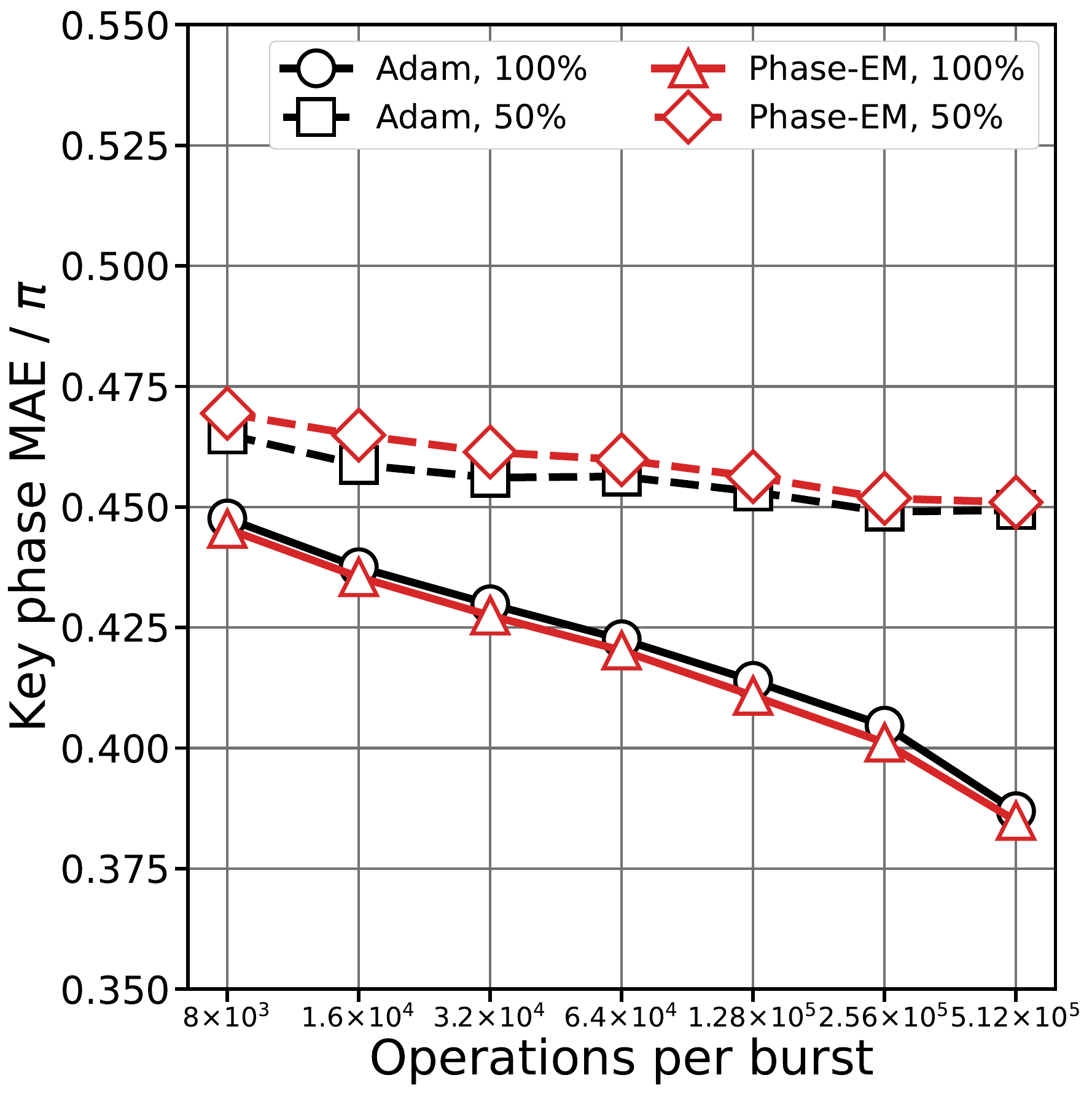}}
    \caption{Online fitting-based detection under the \emph{smooth
    position-dependent key}: balanced detection accuracy (a) without and (b)
    with harmonic-structure initialization; phase-key \gls{mae} (c) without and (d)
    with harmonic-structure initialization.}
    \label{fig:smooth_fitting_detector_validation}
\end{figure*}

\subsubsection{Sequence Detection and Root-Identification Validation}
{\color{black}We next consider a stronger Eve that knows the waveform and
phase-key generation rules. We evaluate offline sequence identification and
online key fitting. The \gls{cnn} represents the offline attack, detecting nominal
sensing sequences and identifying their roots without explicit key recovery.}
The gate and \gls{cnn} are trained on $R=2^{10}=1024$ optimized nominal roots with
\textcolor{black}{$N=64$, $k=5$, $\beta=0.5$ for the protected cases
($\beta=0$ for the unprotected baseline)}, and independently sampled
\gls{snr} over $1$--$10$~dB. The
gate uses 10 bursts from each nominal root and an equal number of random
sequences, while the \gls{cnn} uses 30 root-labeled nominal
bursts per root.
\textcolor{black}{Noise, tangent-\gls{an}, and phase keys are independently
regenerated per burst.} All cases use the same maximum training budget of 60
epochs, with batch sizes of 64 for the gate and 128 for the \gls{cnn}.
Calibration and testing are disjoint, with 100 seen nominal, 100 unseen
nominal, and 100 random test bursts.
With 8-bit quantization per parameter, the \emph{shared} and \emph{smooth} keys
use 40 and 120 bits per burst, respectively; the \emph{seed-expanded} rule
uses a 10-bit seed. Signaling cost therefore increases from
\emph{seed-expanded} to \emph{shared} to \emph{smooth}.

Across all four scenarios, the (a) and (b) panels of
Fig.~\ref{fig:eve_detector_baselines} show the nominal/random \gls{psr}
distributions and the joint \gls{psr}--normalized-sidelobe-energy space with
the learned decision boundary, respectively. These panels
illustrate how well Eve can separate nominal from random bursts. The
distributions are well separated for the unkeyed baseline, enabling reliable
detection. They become increasingly merged under the \emph{shared},
\emph{smooth}, and \emph{seed-expanded} keys, with the last case almost fully
overlapping.
The (c) panels report the resulting gate and root-ID outcomes. Without keying
or \gls{an}, Eve obtains
95\% seen-nominal acceptance, 91\%
unseen-nominal acceptance and random-burst rejection, and 95\% overall
seen-root identification. The \emph{shared-key} case gives 75\%, 72\%, 68\%, and
75\%, whereas the \emph{smooth-key} case gives 67\%, 60\%, 53\%, and 57\%.
Among bursts that pass the gate, the seen-root identification rates are 100\%,
100\%, and 85.1\%, respectively. No wrong root is
accepted in the 100 seen-root samples of any case.
For the \emph{seed-expanded key}, the strongly oscillating curves indicate that,
under the fixed learning budget, neither the gate nor the \gls{cnn} learns stable
discriminative features.

The (d) panels show the maximum \gls{gf} score over shortlisted candidates;
the dashed lines are calibrated thresholds. Scores are separated for the
unkeyed and \emph{shared-key} cases, partly merged for the \emph{smooth key}, and nearly
indistinguishable for the \emph{seed-expanded key}. In the last case, upstream gate
and \gls{cnn} failure also prevents useful calibration.

Relative to the unprotected baseline, all key rules reduce root identification.
The \emph{shared} $k$-phase vector cancels from adjacent-symbol products and centered
Gram features, leaving a stable signature for Eve. The \emph{smooth key} perturbs
this signature across symbols, while the pseudorandom mask expanded from the
shared seed and burst counter removes its reuse and remains difficult to learn
under the fixed budget.
The limited seed space supports tractable evaluation; \textcolor{black}{a larger seed space can further
strengthen key security with modest signaling overhead}. The first two rules
\textcolor{black}{have negligible key-generation cost}, whereas the \emph{seed-expanded rule} requires
the same pseudorandom expansion at both legitimate nodes, trading additional
computation for lower signaling overhead and resistance to
waveform-learning attacks.

\subsubsection{Online Fitting-Based Detection}
We evaluate the Adam and phase-aware \gls{em} detectors in two cases. The
upper-bound case gives Eve the complete nominal sequence but not
the burst key or tangent-\gls{an} realizations. In the partial-knowledge case,
labels are available at 50\% of the positions: 40\% of all labels are correct,
and 10\% are incorrect. Adam excludes unknown positions from its loss, whereas
phase-aware \gls{em} assigns them uniform \glsentryshort{qpsk} priors.
Both methods use a learning rate of $0.03$. At each operation budget, Eve
first fits the candidate waveform and computes its fitting score. A
method-specific threshold is then calibrated on 50 nominal and 50 random
bursts \textcolor{black}{to maximize balanced accuracy and frozen for evaluation: a score below
the threshold declares a nominal waveform; otherwise, the burst is rejected.}
Balanced accuracy equally weights nominal acceptance and random rejection.

Fig.~\ref{fig:fitting_detector_validation}(a) shows that the balanced accuracy
of both methods generally increases with the operation budget and nominal-sequence
knowledge level.
\textcolor{black}{With 50\% knowledge, Adam
converges faster because it optimizes only the key parameters over known
positions, whereas \gls{em} must jointly update the unknown-symbol posteriors and
the key.} With structural initialization in
Fig.~\ref{fig:fitting_detector_validation}(b), the finite-sample acceptance
reaches unity at the smallest tested budget. In
Fig.~\ref{fig:fitting_detector_validation}(c), the phase-key \gls{mae} decreases with
the operation budget and nominal-sequence knowledge. In both
Fig.~\ref{fig:fitting_detector_validation}(c) and (d), phase-aware \gls{em} reaches a
lower \gls{mae} plateau under partial knowledge.
\textcolor{black}{The incorrect labels contaminate
Adam's fitting set, whereas \gls{em} assigns posterior probabilities to the four
\glsentryshort{qpsk} candidates at unknown positions and uses these
probabilities in the key update, reducing the influence of the
incorrect labels.}

Fig.~\ref{fig:smooth_fitting_detector_validation} repeats the evaluation for
the \emph{smooth key} in \eqref{eq:smooth_burst_key}. Unlike the \emph{shared key} with $k$
phases, Eve must estimate $3k$ parameters and regenerate an $N\times k$ phase
mask at every fitting step, resulting in substantially higher computation.
Fig.~\ref{fig:smooth_fitting_detector_validation}(a) and (b) report balanced
accuracy without and with structural initialization, respectively. Structural
initialization does not significantly improve performance, and recovering the
complete position-dependent phase mask remains difficult. At the largest
budget, the key \gls{mae} is approximately $0.39\pi$ with exact nominal knowledge
and $0.45\pi$ with 50\% knowledge.
\textcolor{black}{A larger operation budget may reduce these
errors, but increases the computational burden on a \gls{uav}-mounted Eve.}

\subsection{Validation of Secure Sensing}

\begin{table}[!t]
    \centering
    \caption{Sensing simulation setup for Fig.~\ref{fig:pd_k2_k6_compare}.}
    \label{tab:sensing_simulation_setup}
    \tabletextsize
    \setlength{\tabcolsep}{3pt}
    \renewcommand{\arraystretch}{1.05}
    \begin{tabular}{@{}p{0.48\columnwidth}p{0.48\columnwidth}@{}}
        \toprule[1.2pt]
        \rowcolor{gray!20}
        \textbf{Parameter} & \textbf{Value} \\
        \midrule
        \rowcolor{gray!8}
        Nominal sequence / harmonic dimension & $N=64$ \glsentryshort{qpsk} symbols / $k=5$ \\
        Active subcarriers / \glsentryshort{an} fraction & $320$ / $\beta=0.5$ \\
        \rowcolor{gray!8}
        Phase-key rule / depth & \emph{Smooth position-dependent key} / $2$ rad \\
        Transmit / range \glsentryshort{ifft} sizes & $512$ / $4096$ \\
        \rowcolor{gray!8}
        Subcarrier spacing / sampling rate & $120$ kHz / $61.44$ MS/s \\
        Cyclic prefix / complete symbol duration & $64$ samples / $9.375\,\mu$s \\
        \rowcolor{gray!8}
        Delay / bistatic excess-path bin spacing & $2.03$ ns / $0.61$ m \\
        Useful-signal \glsentryshort{snr} per echo & $-20$ dB per active subcarrier \\
        \rowcolor{gray!8}
        Carrier frequency / echo interval & $3.5$ GHz / $1$ ms \\
        Target-path delays (receive samples) & $12+\delta_0$, $21+\delta_0$, $35+\delta_0$; $\delta_0\sim\mathcal U[-0.5,0.5]$ \\
        \rowcolor{gray!8}
        Relative target-path powers & $0$, $-8$, $-14$ dB \\
        Clutter delays / relative powers & Randomly drawn from $10$--$50$ samples / $-5$, $-10$ dB \\
        \rowcolor{gray!8}
        Path phases / bistatic path-length rate & Independent $\mathcal U[-\pi,\pi)$ per \glsentryshort{mc} run / $1$ m/s approaching \\
        Target Doppler compensation & Both receivers; range walk retained \\
        Eve timing jitter & Independent Gaussian, $2$ ns \glsentryshort{rms} per echo \\
        \rowcolor{gray!8}
        Search-window width & $40$ m \\
        CA-\glsentryshort{cfar} guard / training / stride & $26$ / $32$ cells per side / $13$ bins \\
        \rowcolor{gray!8}
        False-alarm target / lock tolerance & Search-wide $0.05$ / $\pm2$ nearest-delay bins \\
        \bottomrule[1.2pt]
    \end{tabular}
\end{table}

\begin{figure}[!t]
\centering
\subfloat[]{\includegraphics[width=0.473100\columnwidth,height=0.440800\columnwidth]{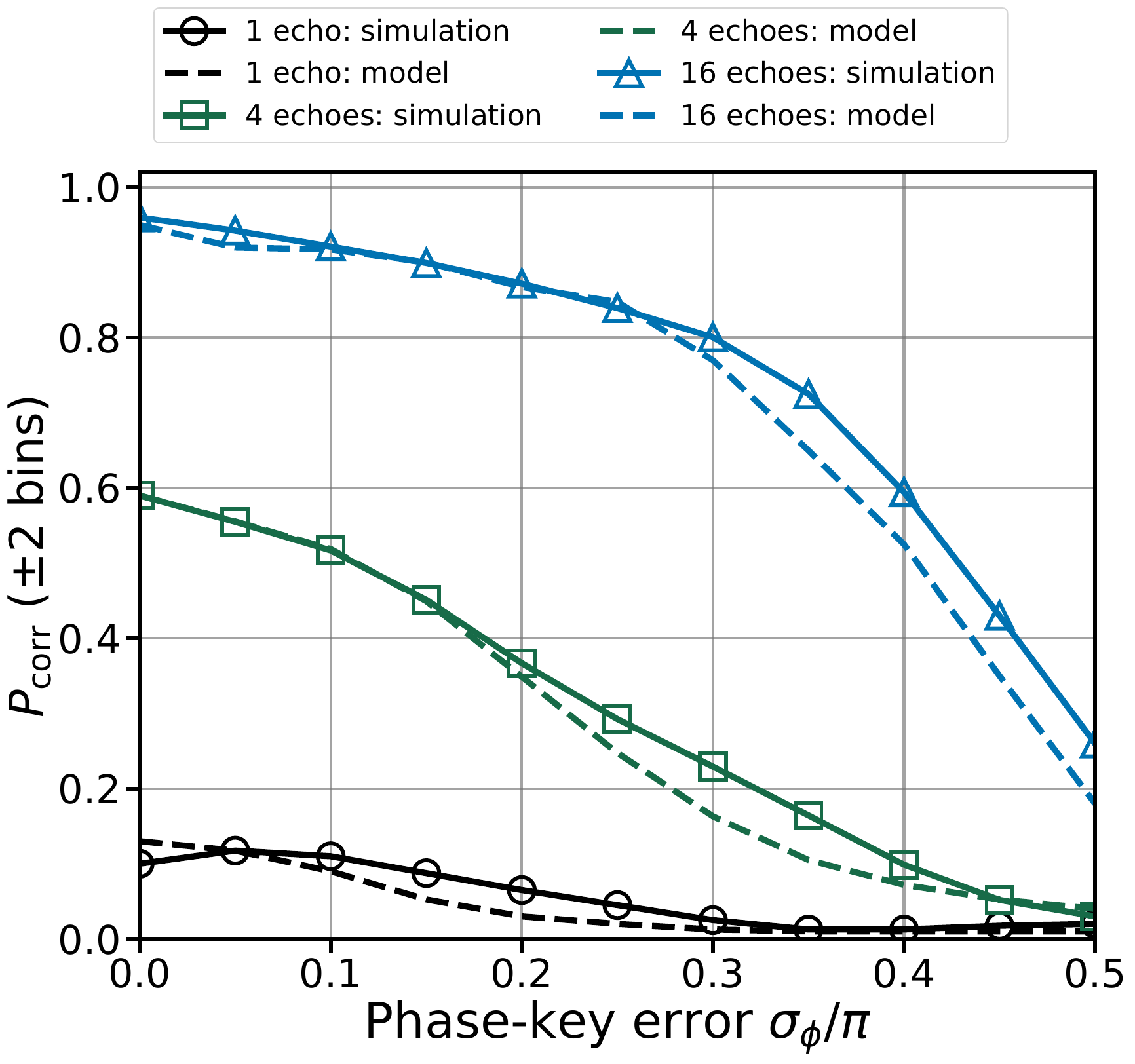}}\hfill
\subfloat[]{\includegraphics[width=0.473100\columnwidth,height=0.440800\columnwidth]{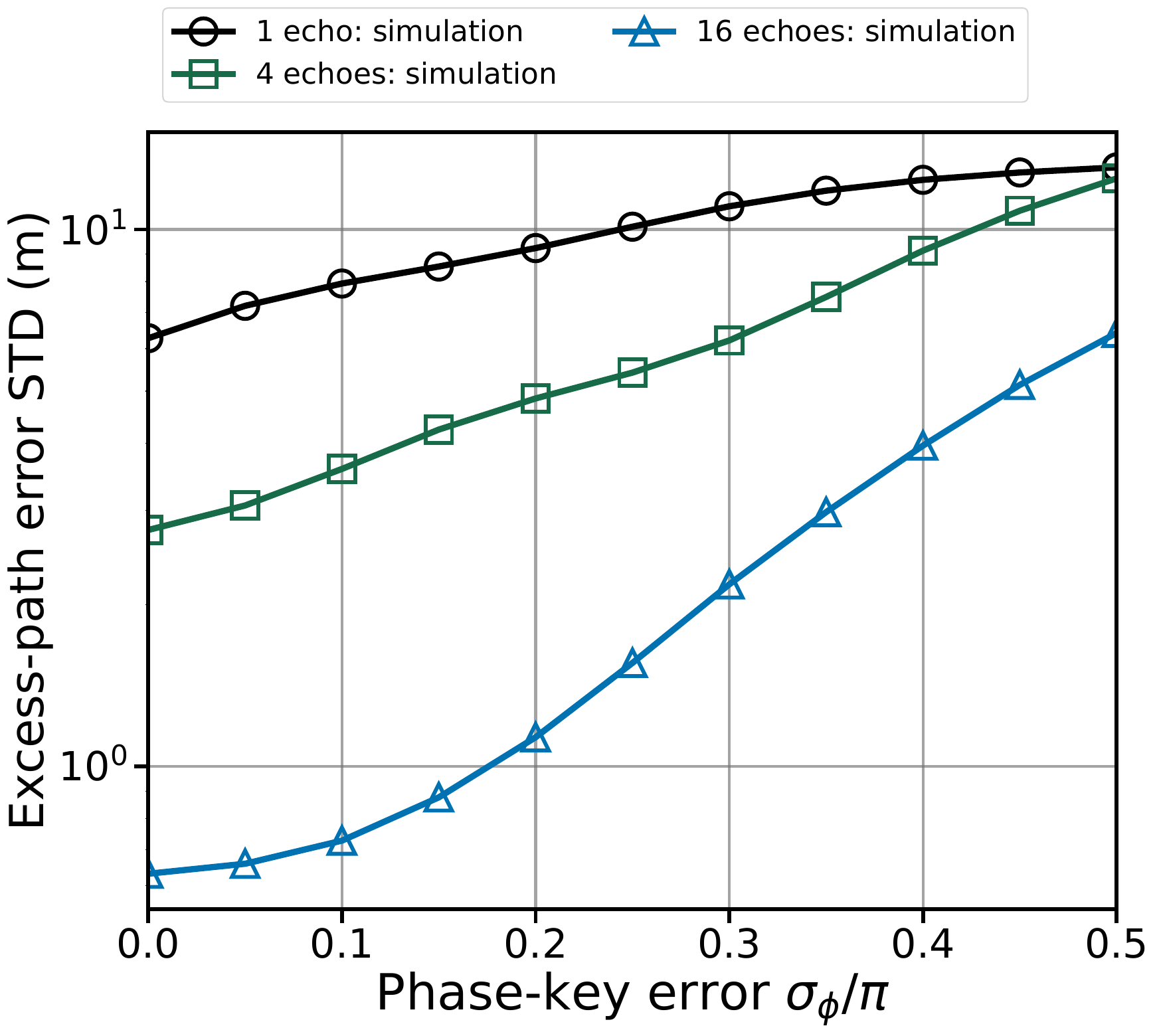}}\par\vspace{1mm}
\subfloat[]{\includegraphics[width=0.473100\columnwidth,height=0.440800\columnwidth]{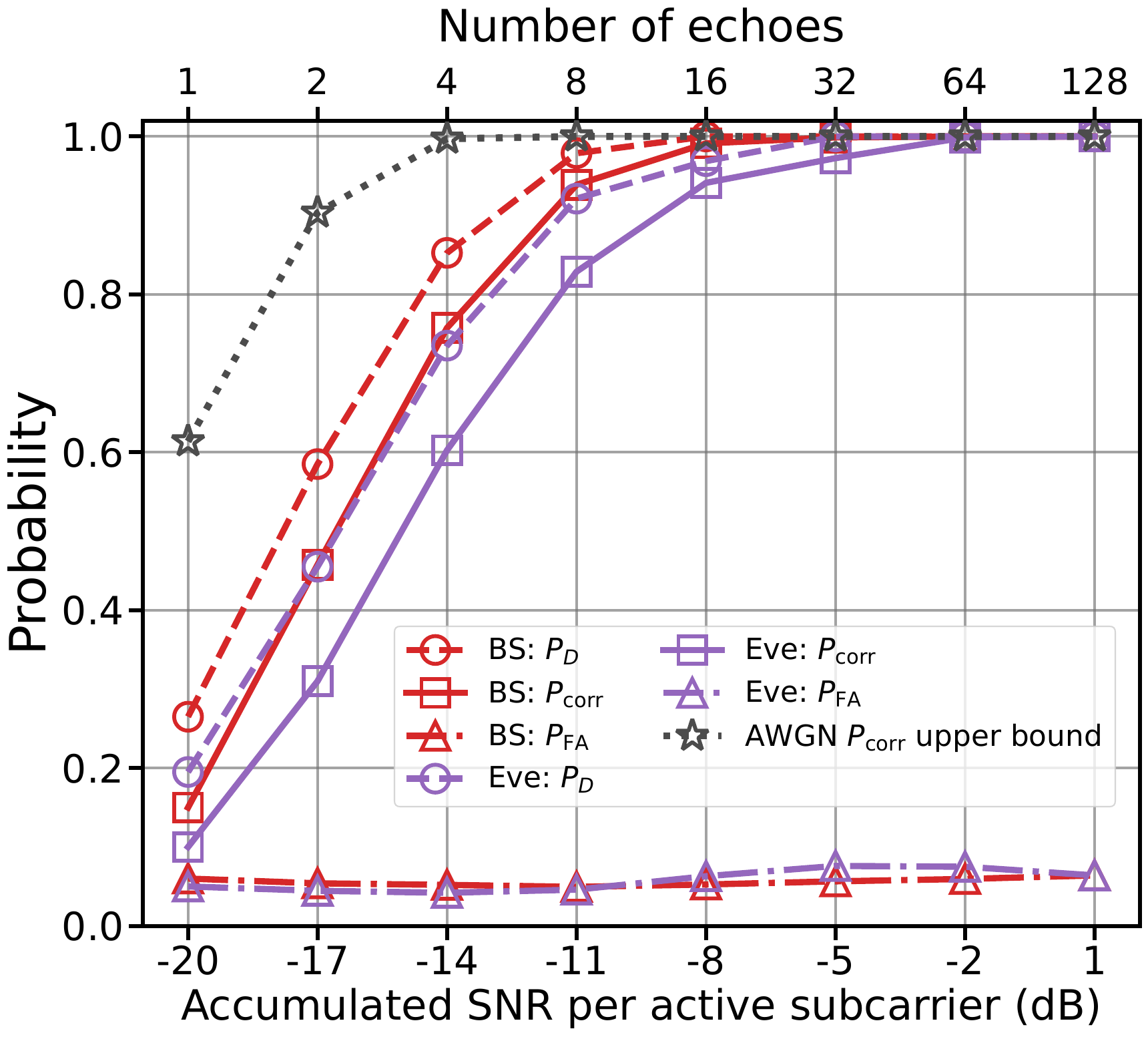}}\hfill
\subfloat[]{\includegraphics[width=0.473100\columnwidth,height=0.440800\columnwidth]{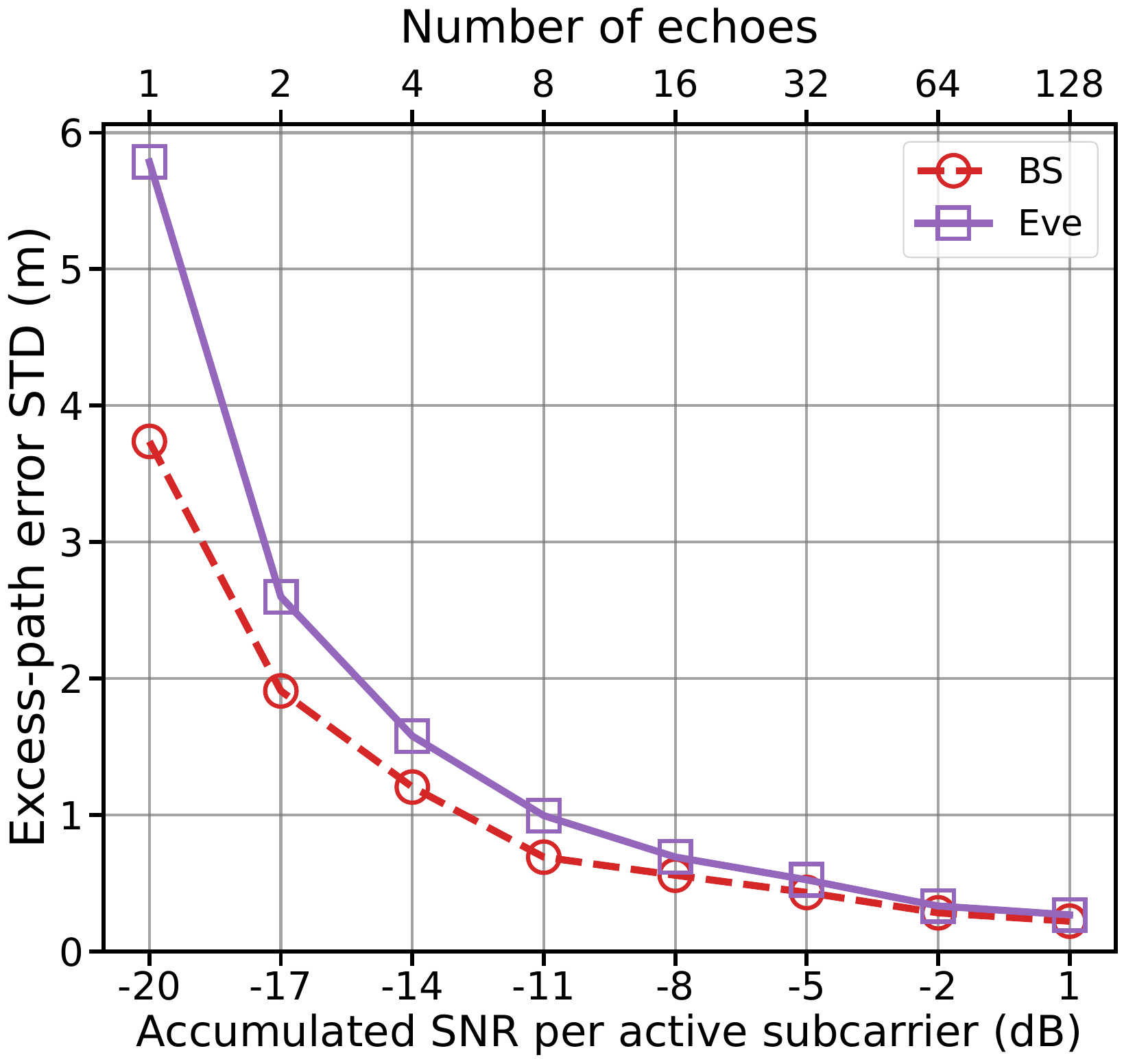}}
\caption{Detection and delay locking, $\beta=0.5$:
(a) $P_{\mathrm{corr}}$ versus key error for $1$, $4$, and $16$ echoes;
(b) corresponding simulated excess-path error \glsentryshort{std};
(c) detection probabilities and \glsentryshort{awgn} $P_{\mathrm{corr}}$ upper bound versus accumulation;
(d) corresponding excess-path error \glsentryshort{std}.
In (c)--(d), Eve: $\sigma_\phi=0.1\pi$.}
\label{fig:pd_k2_k6_compare}
\end{figure}

Here, we evaluate Eve's coherent sensing attack with imperfect phase-key
knowledge, assuming perfect nominal-sequence knowledge and coarse burst
localization. Eve is granted the same beamforming gain as the legitimate
receiver, representing favorable interception geometry. Eve is also assumed to have representative background
clutter-and-noise observations for \gls{cfar} calibration.
Table~\ref{tab:sensing_simulation_setup} summarizes the setup for
Fig.~\ref{fig:pd_k2_k6_compare}.

Fig.~\ref{fig:pd_k2_k6_compare}(a) compares the model predictions with
waveform simulations using $1$, $4$, or $16$ echoes. The analytical model in
\eqref{eq:key_error_noncentrality} and \eqref{eq:key_error_pcorr_model}
converts waveform distortion into effective \gls{snr} degradation to
approximate $P_{\mathrm{corr}}$. Increasing key error reduces correct locking
through two coupled effects: residual phase mismatch weakens coherent
correlation, while the erroneous tangent direction leaves uncancelled
\gls{an}. Accumulating more echoes
improves correct locking, but substantial key errors still degrade performance.
Fig.~\ref{fig:pd_k2_k6_compare}(b) shows the corresponding simulated
excess-path error \gls{std}, including incorrect delay locks. With small key
errors, $16$ echoes maintain submeter \gls{std}. As mismatch grows, the coherent
peak weakens and incorrect delay bins are selected more frequently,
increasing the error spread. More echoes therefore improve both locking
probability and localization precision, but do not eliminate sensitivity
to residual key and \gls{an} errors.

Fig.~\ref{fig:pd_k2_k6_compare}(c)--(d) compares detection and excess-path
error under coherent echo accumulation at an ideal base station and Eve.
The base station perfectly dekeys the waveform and removes the
\gls{an}; Eve estimates \gls{an} using the erroneous key.
The nominal reference is known to both receivers.
The top axes give the echo count $B$; the bottom axes use the common
nominal accumulation scale $-20+10\log_{10}B$ dB.
Eve's delay locking improves more slowly with accumulation,
although independent key and \gls{an} errors can partly average out.
\textcolor{black}{Achieving the assumed key accuracy nevertheless requires prior information
and processing.} In the coherent sensing attack, recovering the \emph{shared
lookup-table key} relies on partial nominal-sequence knowledge, while the
\emph{smooth position-dependent key} additionally entails substantially
higher fitting cost. For the \emph{seed-expanded pseudorandom key}, a known
nominal root leaves $2^\lambda$ key-seed candidates; without root identification,
joint exhaustive search considers $R2^\lambda$ root--seed pairs, where $R$
is the number of candidate roots. Root identification narrows this search
without revealing the seed.

\section{Discussion and Conclusion}

In this work, we investigated waveform-level security for \gls{isac}
against a \gls{uav}-mounted Eve. We combined Fourier-curve modulation,
phase keys, and tangent \gls{an} to protect the sensing reference while
preserving legitimate communication and sensing. To assess protection under
different levels of adversarial knowledge, we considered keyless reception,
offline sequence identification, online key fitting, and the coherent sensing
attack. We evaluated three key-generation rules and related residual
phase-key error and imperfect \gls{an} removal to correct-delay locking.
The numerical results demonstrate reliable legitimate reception and reduced
Eve inference performance. Shared keys retain exploitable structure,
smooth position-dependent keys increase fitting difficulty, and seed-expanded
pseudorandom keys remain difficult for the evaluated detector to learn
under the fixed budget.

The proposed framework supports waveform-level protection beyond spatial
beamforming, while identifying its limits. These findings
motivate joint waveform and spatial protection, long-observation attack
evaluation, and key-recovery latency and energy measurements for constrained
adversaries.

\bibliographystyle{IEEEtran}
\bibliography{references}

\end{document}